\documentclass[12pt]{amsart}
\usepackage{amssymb}
\usepackage{graphicx}
\numberwithin{equation}{section}
\newcommand{\smallpagebreak}{\par\vspace{2 mm}\noindent}
\newcommand{\medpagebreak}{\par\vspace{4 mm}\noindent}
\newcommand{\bigpagebreak}{\par\vspace{6 mm}\noindent}

\newcommand{\dsize}{\textstyle}
\newcommand{\tsize}{\scriptstyle}

\newcommand{\demo}{\smallpagebreak\noindent{\it Proof.\/} \ }

\newcommand{\point}[1]{\par\vspace{1mm}({\it #1\/}) \ }

\newcommand{\N}{{\mathbb N}}
\newcommand{\R}{{\mathbb R}}
\newcommand{\Z}{{\mathbb Z}}

\newcommand{\C}{{\mathbb C}}
\newcommand{\MI}{{\mathbb M}}
\newcommand{\IK}{{\mathbb K}}
\newcommand{\HI}{{\mathbb H}}

\newcommand{\im}{{\rm Im}\,}
\newcommand{\re}{{\rm Re}\,}

\newcommand{\ctg}{{\rm ctg}\,}
\newcommand{\Const}{{\rm Const}\,}

\newcommand{\Res}{{\rm Res}\, }
\newcommand{\cc}[1]{\overline #1}

\title{On the difference equations with periodic
coefficients}
\author{  V.Buslaev and  A.Fedotov}
\address{ Department of Mathematics and Mathematical Physics,
  St.Petersburg State University,  1, Ulianovskaja, St.Petersburg-Petrodvorets
 Russia}
\email{buslaev@mph.phys.spbu.ru, fedotov@mph.phys.spbu.ru}
\begin{document}
\begin{abstract}
In the paper, we study entire solutions of the difference equation 
$\psi\,(z+h)=M\,(z)\,\psi\,(z)$, \ $z\in\C$, \ $\psi\,(z)\in\C^2$. 
In this equation, $h$ is a fixed positive parameter, and $M:\C\mapsto
SL\,(2,\,\C)$ is a given matrix function. We assume that $M(z)$ is a 
$2\pi$-periodic trigonometric polynomial. The main aim is to construct 
the minimal entire solutions, e.i. the solutions with the minimal possible 
growth simultaneously as for $z\to -i\infty$ so for $z\to+i\infty$. \\ 
We show that the monodromy matrices corresponding to the bases made of the 
minimal solutions are trigonometric polynomials of the same order as the 
matrix $M$. This property relates  the spectral analysis of the one 
dimensional difference Schr\"odinger equations with the potentials being 
trigonometric polynomials to an analysis of a finite dimensinal dynamical 
system. 
\end{abstract}
\maketitle
\section{Introduction: Some definitions and main
results} The paper is devoted to the study of entire solutions of
the equation
\begin{equation}\psi\,(z+h)=M\,(z)\,\psi\,(z),\quad z\in\C,\quad
 \psi\,(z)\in\C^2. \label{1.1}\end{equation}
In this equation, $h$ is a fixed positive parameter, and $M$ is a
given matrix. We assume that the matrix $M$ satisfies the
following two conditions. First, $$M\,(z)\in SL\,(2,\,\C),\quad
z\in\C,$$ secondly, the matrix $M$ is a $2\pi$-periodic
trigonometric polynomial.
\smallpagebreak
Since equation
\eqref{1.1} is invariant with respect to the multiplication of the
solutions by the $h$-periodic functions, one can construct its
entire solutions growing as for  $z\to-i\infty$ so for
$z\to+i\infty$ as quickly as wanted. The main aim of this paper is
to construct the minimal entire solutions, e.i. the solutions with
the minimal possible growth simultaneously as for $z\to -i\infty$
so for $z\to+i\infty$. 
\smallpagebreak
The set of solutions of \eqref{1.1} is
invariant with respect to the operator of $2\pi$-translation:
$f(z)\mapsto f(z+2\pi)$, and there are natural objects, monodromy
matrices, describing the transformations of the bases in the space
of solutions of \eqref{1.1} under such translations, see
subsection \ref{mat-mon}. 
The notion of the monodromy matrix 
for difference equations with periodic coefficients is a natural
generalization of the classical notion of the monodromy matrix 
for ordinary differential equations with periodic coefficients. 
For the differential equations, the monodromy matirces are 
constant, i.e. independent of the variable of the equation.
For the difference equations, the monodromy matrices are
periodic functions of the variable. We show that the monodromy matrices
corresponding to the bases made of the minimal solutions have the
most simple analytic structure: they are trigonometric
polynomials of the same order as the matrix $M$.
\smallpagebreak
The monodromy matrices appear in  the spectral analysis
of one dimensinal  Schr\"odinger equations with periodic potentials.
The spectral analysis of a differential Schr\"odinger
equation with a periodic potential reduces to the  study of the spectrum 
of a (constant) monodromy matrix. This leads a simple ``discret'' band 
structure of the spectra of the periodic differential equations. 
For difference equations, the monodromy matrices being periodic, instead of 
the spectral analysis of an individual monodromy matrix, one arrives to an 
infinite sequence of the monodromy matrices and has to study properties
of an infinite sequence of finite difference equations of the form~\eqref{1.1}.
This reflects the cantorien structure of the spectra of 
difference equations with periodic coefficients. The above 
property of the monodromy matrices corresponding to the minimal entire 
solutions is important for the spectral
analysis of the  difference Schr\"odinger equations
with the potentials being trigonometric polynomials.
It relates their spectral analysis to an
analysis of a finite dimensinal dynamical system.
\smallpagebreak 
This paper is a natural
continuation of the articles \cite{BF1}, \cite{BF2}, 
\cite{BF3}, \cite{BF4}, \cite{BF5}, \cite{BF6} and \cite{BF7} devoted to Harper equation 
\begin{equation}
\frac{\psi\,(z+h)+\psi\,(z-h)}2+\cos
z\,\psi\,(z)=E\psi\,(z)\label{1.2}\end{equation} 
and was inspirated by  the papers of B.Helffer 
and J.Sj\"ostrand, and of Wilkinson (see, for example, \cite{HS}, \cite{W}).
They have suggested an asymptotic renormalization method 
to study the spectrum of the Harper equation~\eqref{1.2}.
Our works are related to an exact renormalization procedure, 
the monodromization method.
\subsection{Notations and agreements} \ {\bf 1.} \
To simplify the notations, we let $$a\,(z)=M_{11}(z),\quad
  b\,(z)=M_{12}(z),\quad
  c\,(z)=M_{21}(z),\quad
  d\,(z)=M_{22}(z).$$

\smallpagebreak {\bf 2.} \ If either $b\,(z)\equiv0$, or
$c\,(z)\equiv0$, then the vector equation \eqref{1.1} reduces to
two scalar equations which can be solved explicitly. In this
paper, we always assume that \begin{equation} b\,(z)\not\equiv
0,\label{1.3}\end{equation} and let
\begin{equation} \rho\,(z)={b\,(z)}/b\,(z-h),\quad\quad
  v\,(z)=a(z)+\rho\,(z)\,d\,(z-h).\label{1.4}\end{equation}
These two functions play an essential part in our constructions.
Elementary calculations show that if $\psi$ is a vector solution
of \eqref{1.1}, then its first component satisfies the equation
\begin{equation}
\psi_1(z+h)+\rho\,(z)\,\psi_1(z-h)=v\,(z)\,\psi_1(z).\label{1.5}
\end{equation}

\smallpagebreak {\bf 3.} \ When describing the  vector solutions
of \eqref{1.1}, we describe only their first components: if the
first component of a vector solution is known, the second one can
be reconstructed by the formula
\begin{equation} \psi_2(z)=\frac{\psi_1(z+h)-a\,(z)\,\psi_1(z)}{b\,(z)}.\label{1.6}\end{equation}

\smallpagebreak
{\bf 4.} \  Throughout the paper, we use also the following notations.
Let $f$ be a $2\pi$-periodic function. If
$$f\,(z)=f_{-m} e^{-imz}(1+o\,(1)),\quad f_{-m}\ne0,\quad z\to+i\infty,$$
we put
$$n_+(f)=m,\quad f_+=f_{-m},$$
and if
$$f\,(z)=f_{l} e^{ilz}(1+o\,(1)),\quad f_{l}\ne0,\quad z\to-i\infty,$$
we let
$$n_-(f)=l,\quad f_-=f_{l}.$$
\subsection{Set of analytic solutions of
\eqref{1.1}}  List some elementary properties of the set of
analytic solutions of equation \eqref{1.1}, see, for example,
\cite{BF7}.

\smallpagebreak Fix $\alpha,\,\beta\in\R$, \ $\alpha<\beta$. Let
$S$ be the strip $$S=\left\{z\in\C:\,\,\alpha<\im
z<\beta\right\}$$ and let $\MI\,(S)$ be the set of the vector
solutions of \eqref{1.1} analytic in this strip. Denote the ring
of all $h$-periodic functions analytic in $S$ by $\IK\,(S)$.

\smallpagebreak For any two solutions $\psi_1,\,\psi_2\in
\MI\,(S)$ the wronskian
\begin{equation} \{\psi_1\,(z),\,\psi_2\,(z)\}\equiv\det\left(\psi_1(z),\,\psi_2(z)\right)
\label{1.7}\end{equation} is an element of $\IK\,(S)$. The
solutions $\psi_1$, $\psi_2$ are linearly independent over the
ring $\IK\,(S)$  if $\{\psi_1(z),\psi_2(z)\}\ne 0$, $z\in S$.

\smallpagebreak Let two solutions $\psi_1,\,\psi_2\in \MI\,(S)$ be
linearly independent over the ring $\IK\,(S)$. Then any solution
$\psi\in\MI\,(S)$ can be uniquely represented in the form
\begin{equation} \psi=\alpha\,\psi_1+\beta\,\psi_2,\quad \alpha,\beta\in
\IK\,(S).\label{1.8}\end{equation}
Thus, the set $\MI\,(S)$ is a two dimensional module over the ring
$\IK\,(S)$, and  the solutions $\psi_1,\,\psi_2$ form a basis of
$\MI\,(S)$.

\smallpagebreak In the case of $S=\C$, we write simply  $\MI$ and
$\IK$.
\subsection{Bloch solutions} The set $\MI\,(S)$ is
invariant with respect to the $2\pi$-translations. We call $f\in
\MI\,(S)$ a Bloch solution if
\begin{equation} f\,(z+2\pi)=u\,(z)\,f\,(z),\quad u\in \IK\,(S).
\label{1.9}\end{equation}
The factor $u$ is called the Bloch multiplier of the solution $f$.
Bloch solutions play an important role in the theory of
\eqref{1.1}.
\subsection{ Singular points of equation
\eqref{1.1}}  The $2\pi$-periodicity of the matrix $M$ makes
natural to consider $+i\infty$ and $-i\infty$ as two singular
"points" of the equation. For any fixed $Y\in\R$, call the
half-plane $\C_+=\C_+(Y)=\{z\in\C\,:\im z>Y\}$ a vicinity of
$+i\infty$, and the half-plane $\C_-=\C_-(Y)=\{z\in\C\,:\im z<Y\}$
a vicinity of the $-i\infty$. For brevity, we shall write $$
\MI_\pm=\MI\,(\C_\pm),\quad \IK_\pm=\IK\,(\C_\pm).$$

\smallpagebreak The {\it minimal} entire solutions of equation
\eqref{1.1} can be characterized as the entire solutions having
the "simplest" asymptotic behavior as for $z\to-i\infty$ so for
$z\to+i\infty$.
\subsection{ Canonical bases in vicinities of the
singular points} We shall define the minimal  solutions in terms
of the "simplest" solutions living and analytic in some vicinities
of the singular points $\pm i \infty$. They appear to be Bloch
solutions. One has

\medpagebreak {\bf Theorem 1.1a.} \ {\it Let $b\not\equiv0$, and
let $v\to\infty$ as $z\to+i\infty$. There exist two vector Bloch
solutions $f_{1,2}$ of \eqref{1.1} analytic in   a vicinity $\C_+$
of $+i\infty$  with the first components admitting the
representations \begin{equation} {(f_{1,2})}_1(z)=e^{\dsize
\pm\frac{i}{2hn_+(v)}\,(n_+(v)\,z+\phi_+)^2+
i(n_+(v)-n_+(b))\,\frac{z}2+o\,(1)},\label{1.10}\end{equation} as
$z\to+i\infty$.  Here, \begin{equation} \phi_+=\dsize i\ln
v_+-\frac{h}2\,n_+(b).\label{1.11}\end{equation} These two
solutions are linearly independent over the ring $\IK_+$,
\begin{equation} \det\left(f_1(z),\,f_2(z)\right)=\frac{v_+}{b_+}.\label{1.12}\end{equation}
}

\medpagebreak {\bf Theorem 1.1b.} \ {\it Let $b\not\equiv0$, and
let $v\to\infty$ as $z\to-i\infty$. There exist two Bloch
solutions of \eqref{1.1} analytic in   a vicinity $\C_-$ of
$-i\infty$ with the first components admitting the representations
\begin{equation} {(g_{1,2})}_1(z)=e^{\dsize
\pm\frac{i}{2hn_-(v)}\,(n_-(v)z+\phi_-)^2-
i(n_-(v)-n_-(b))\,\frac{z}2+o\,(1)},\label{1.13}\end{equation} as
$z\to-i\infty$. Here, \begin{equation} \phi_-=-i\ln
v_--\dsize\frac{h}2\,n_-(b).\label{1.14}\end{equation} These two
solutions are linearly independent over the ring $\IK_-$,
\begin{equation} \det\left(g_1(z),\,g_2(z)\right)=-\frac{v_-}{b_-}.\label{1.15}\end{equation}
}

\medpagebreak
{\bf Remark 1.} \ {\small Note that the numbers $\phi_\pm$ are defined
only modulo $2\pi$. Fixing, for example,  the parameter $\phi_+$
in two different ways, we obtain two different pairs $f_{1,2}$.}

\smallpagebreak {\bf Remark 2.} \ {\small The asymptotic formulas
for $f_{1,2}$ and $g_\pm$ immediately imply the asymptotics of the
corresponding Bloch multipliers. For $f_{1,2}$ the Bloch
multipliers admit the representations
\begin{equation} \alpha_{1,2}(z)=\alpha_{1,2}^0\,e^{\dsize \pm 2\pi n_+(v) iz/h
}(1+o\,(1)), \quad z\to+i\infty,\label{1.16}\end{equation} and,
for $g_{1,2}$, the Bloch multipliers have the form
\begin{equation} \beta_{1,2}(z)=\beta_{1,2}^0\,e^{\dsize \pm 2\pi n_-(v) iz/h
}(1+o\,(1)), \quad z\to+i\infty,\label{1.17}\end{equation} here
$\alpha_{1,2}^0$ and $\beta_{1,2}^0$ are nonzero constants,
$$\alpha_{1,2}^0= \exp\left(\pm\frac{2\pi i}h\,(\phi_++\pi
n_+(v))+i\pi(n_+(v)-n_+(b)) \right),$$ $$\beta_{1,2}^0=
\exp\left(\pm\frac{2\pi i}h\,(\phi_-+\pi
n_-(v))-i\pi(n_-(v)-n_-(b)) \right).$$ }

\medpagebreak
We see that the  pair $f_{1,2}$ is a basis of $\MI_+$, and the solutions
$g_{1,2}$ form a basis of $\MI_-$. We call these bases
{\it canonical}. We call the number $\phi_+$ {\it the parameter
of the canonical basis $f_{1,2}$} and the number $\phi_-$
{\it the parameter of the canonical basis $g_{1,2}$}.
\subsection{Consistent canonical bases} Let
$$\Sigma=\{\pm (2\pi+h +2hl+2\pi m),\,\,\,l,m\in\N\}.$$ We call
two canonical bases $f_{1,2}$ and $g_{1,2}$ {\it consistent} if
their parameters satisfy the condition
\begin{equation} \phi_+-\phi_-\not\in\Sigma.\label{1.18}\end{equation} Since the
numbers $\phi_\pm$ are defined  modulo $2\pi$, the consistent
bases always exist.
\subsection{Minimal entire solutions. Existence} Let
the assumptions of Theorems 1a -- 1b be fulfilled. Fix two
canonical bases $f_{1,2}$ and $g_{1,2}$ analytic in some
vicinities $\C_\pm$ of $\pm i\infty$. Any entire solution of
\eqref{1.1} admits the representations
\begin{equation} \psi\,(z)=A\,(z)\,f_1(z)+B\,(z)\,f_2(z),\quad z\in \C_+,\quad
A,\,B\,\in \IK_+,\label{1.19}\end{equation}
\begin{equation}
\psi\,(z)=C\,(z)\,g_1(z)+D\,(z)\,g_2(z),\quad z\in C_-,\quad
C,\,D\,\in \IK_-.\label{1.19a}\end{equation}
 Note that the periodic coefficients in these
representations are uniquely defined.

\smallpagebreak We call a nonzero entire solution of \eqref{1.1}
{\it minimal}, if \point{ \ i} its coefficients $A$, $B$ are
bounded in $\C_+$; \point{ ii}  the coefficients $C$ and $D$ are
bounded in $\C_-$; \point{iii} at least, one of these four
coefficients tends to zero as $\im z$ tends to the corresponding
infinity.

\noindent
Note that this definition depends on the choice
of the canonical bases. For a given pair of canonical bases, the definition
singles out at least four minimal solutions.

\smallpagebreak
One of the main results of the paper is

\medpagebreak {\bf Theorem 1.2.} \  {\it Let $b\,(z)\not\equiv 0$,
and \begin{equation} n_+(v)=n_-(v)=n>0.\label{1.20}\end{equation}
Then, for any two consistent canonical bases, there exist all the
four corresponding minimal entire solutions.}

\medpagebreak Note that under the assumptions of this theorem,
there exist all the canonical bases. We have already explained the
condition on the coefficient $b$. Discuss the assumption
\eqref{1.20}. We shall see that among the equations \eqref{1.1},
there is a nontrivial one which can be solved in terms of certain
contour integrals. This is the equation with \begin{equation}
a\,(z)=-2\lambda\,\cos(nz),\quad b\,(z)=-1,\quad c\,(z)=1,\quad
d\,(z)=0, \label{1.21}\end{equation} where $n\in \N$, and
$\lambda$ is a complex number. The proof of the existence theorem
is based on this observation. In the case of \eqref{1.21}, \
$v\,(z)=2\lambda\cos(nz)$. This leads to \eqref{1.20}.

\medpagebreak
Now, we  shall discuss  the basic properties of the minimal solutions.
\subsection{Asymptotic coefficients of the minimal
solutions} Consider the representations \eqref{1.19} for a minimal
solution. The coefficients $A$, $B$, $C$ and $D$ can be
represented by the converging Fourier series
$$A\,(z)=\sum_{l=0}^\infty A_le^{\dsize 2\pi ilz/h},\quad
B\,(z)=\sum_{l=0}^\infty B_le^{\dsize 2\pi ilz/h},\quad   z\in
\C_+,$$ $$C\,(z)=\sum_{l=0}^\infty C_le^{\dsize -2\pi ilz/h},\quad
D\,(z)=\sum_{l=0}^\infty D_le^{\dsize - 2\pi ilz/h},\quad  z\in
\C_-.$$ One of the coefficients $A_0$, $B_0$, $C_0$ and $D_0$ is
zero.

\smallpagebreak
Assume that the Fourier coefficient
$A_0$  equals to $0$. In this case,
we denote the minimal solution  by $\psi_A$, and call the Fourier
coefficients
$$A_1,\quad
B_0,\quad
C_0 \quad {\rm and}\quad D_0$$
the asymptotics coefficients of this minimal solution.

\smallpagebreak
If $A_0\ne0$, but $B_0=0$,  we denote the minimal solution by
$\psi_B$, and call the Fourier coefficients
$$A_0,\quad B_1,\quad C_0 \quad {\rm and}\quad D_0$$
the asymptotics coefficients.

\smallpagebreak
Continuing in the same way, we define the minimal solutions
$\psi_C$ and $\psi_D$, and their asymptotic coefficients.

\smallpagebreak
In the sequel, we denote the asymptotic coefficients of the minimal
solution $\psi_A$ by ${A_A}$, \ ${B_A}$, \ ${C_A}$, \ ${D_A}$,
the asymptotic coefficients of
the minimal solution $\psi_B$ by  ${A_B}$, \ ${B_B}$, \ ${C_B}$, \ ${D_B}$
and so on.

\smallpagebreak
The asymptotic coefficients play a crucial part
in the analysis of the minimal entire solutions.
Let us describe  some elementary observations.
\subsection{ Wronskians of the minimal solutions}
Let there exist two canonical bases $f_{1,2}$ and $g_{1,2}$ in
some vicinities of $\pm i\infty$. Denote the wronskian of
$f_{1,2}$ by $w_f$ and let $w_g$ be the wronskian of $g_{1,2}$.
Assume also that there exist all the corresponding minimal
solutions. In Section 6.1, we prove

\medpagebreak
{\bf Proposition 1.3.} \ {\it The wronskians of any two of the  minimal
solutions
are constant. One has
$$\{\psi_A,\,\psi_B\}=-w_f\,{B_A}\,{A_B},$$
$$\{\psi_A,\,\psi_C\}=-w_f\,{B_A}\,{A_C}=w_g\,{C_A}\,{D_C},\quad\quad
\{\psi_A,\,\psi_D\}=-w_f\,{B_A}\,{A_D}=-w_g\,{D_A}\,{C_D},$$
$$\{\psi_B,\,\psi_C\}=w_f\,{A_B}\,{B_C}=w_g\,{C_B}\,{D_C},\quad\quad
\{\psi_B,\,\psi_D\}=w_f\,{A_B}\,{B_D}=-w_g\,{D_B}\,{C_D},$$
$$\{\psi_C,\,\psi_D\}=-w_g\,{D_C}\,{C_D}.$$}

\medpagebreak
This statement immediately implies

\medpagebreak {\bf Corollary 1.4.} \ {\it If all the asymptotic
coefficients are nonzero, then any two of the minimal solutions
form a basis of $\MI$.}

\medpagebreak
In Section 6.2, we obtain also

\medpagebreak {\bf Corollary 1.5. (Uniqueness Theorem)} \ {\it If
all the asymptotic coefficients are nonzero, then any of the
minimal solutions is unique up to a constant factor.}

\medpagebreak The existence theorem is the central result of this
paper. Remind that the matrix $M$ is a trigonometric polynomial.
In the next publication, we shall study a family of the equations
\eqref{1.1} parametrized by the constant coefficients of $M$. In
particular, we shall prove that the minimal solutions and their
asymptotic coefficients can be considered as meromorphic functions
of these parameters, and that these functions are not identically
zero. Thus, we shall see that the asymptotic coefficients are non
zero for any {\it typical} matrix $M$.
\subsection{Monodromy matrices} \label{mat-mon} The
notion of the minimal solution is the first of the main notions of
this paper, and the second one is the notion of a monodromy
matrix. We begin by recalling the general definition of a
monodromy matrix and the description of the monodromization
procedure, see \cite{BF5}.

\smallpagebreak Now, we do not need to assume that the matrix $M$
is a trigonometric polynomial. In fact, we have to suppose only
that it belongs to $SL\,(2,\,\C)$ and is a $2\pi$-periodic matrix
function.

\medpagebreak {\bf 1.} \ Let $\Psi$ is a matrix solution of
\eqref{1.1}.  We call this solution {\it fundamental} if
$\det\Psi\,(z)\equiv\Const\ne0$. Note that $\Psi(z+2\pi)$ is a
solution of \eqref{1.1} together with $\Psi(z)$. We define the
{\it monodromy matrix} corresponding to a given fundamental
solution $\Psi$ by the relation
$$\Psi\,(z+2\pi)=\Psi\,(z)\,M_1^t(z),$$ where ${}^t$ is the
transposition. The function $M_1$ has the properties:
\begin{equation} \det M_1(z)\equiv1,\quad
M_1(z+h)=M_1(z).\label{1.22}\end{equation}

\smallpagebreak Let $\psi^{(1)}$ and $\psi^{(2)}$ be two vector
solutions of \eqref{1.1}. Compose of them the matrix
$\left(\psi^{(1)},\,\psi^{(2)}\right)$. If its determinant is a
nonzero constant, then this matrix is a fundamental solution. We
define the monodromy matrix for such $\psi_1$ and $\psi_2$ as the
monodromy matrix corresponding to the fundamental solution
$\left(\psi^{(1)},\,\psi^{(2)}\right)$.

\smallpagebreak The notion of a monodromy matrix is well known in
the theory of the ordinary differential equations with periodic
coefficients.  For a differential equation
$$\Psi'(z)=M\,(z)\,\Psi\,(z)$$ with a $2\pi$-periodic matrix $M$,
the monodromy matrix also defined by
$\Psi\,(z+2\pi)=\Psi\,(z)\,M_1^t$, but, it is independent of $z$.

\medpagebreak {\bf 2.} \ Recall the description of the
monodromization procedure. Consider the sequence of the numbers
$h_j$, \ $j=0,1,\dots$, defined by the relations $$h_0=2\pi,\quad
h_1=h,$$ $$h_{j-1}=p_j h_j+h_{j+1},\quad 0\le h_{j+1}<h_j,\quad
p_j\in \N.$$ The  $p_j$ are the denominators of the continued
fraction $$h_1/h_0= \frac 1{\dsize p_1+\frac 1{\dsize p_2+\frac
1{\dsize p_3+\dots}}}, $$

\smallpagebreak Let $\Psi_0$ be a fundamental solution of equation
\eqref{1.1}, and let $M_1$ be the corresponding monodromy matrix.
Bring into the consideration the equation
\begin{equation} \Psi_1(z+h_2)=M_1(z)\,\Psi_1(z).\label{1.23}\end{equation} In
view of \eqref{1.22}, this equation is of the same type as
\eqref{1.1}.

\smallpagebreak The passage from \eqref{1.1} to \eqref{1.23} is
the first step of the monodromization procedure. If there exists a
fundamental solution of \eqref{1.23}, the {\it monodromization
procedure} can be continued. In result, one obtains the suite of
the monodromy matrices satisfying the relations: $$\det
M_j(z)\equiv1,\quad M_j(z+h_j)=M_j(z).$$

\smallpagebreak
If the number $h_1/h_0$ is rational, the procedure is finite,
In general case, the {\it monodromization procedure\/} is infinite.

\smallpagebreak 
The spectral analysis of a differential 
equation with periodic coefficients reduces to the  study the spectrum 
of a (constant) monodromy matrix. 
Trying develop similar ideas  for the difference equations,  arrives 
to the infinite sequence of finite difference equations~\eqref{1.22}.
In \cite{BF4}--\cite{BF5}, we have seen that
spectral properties of \eqref{1.1} with respect to the parameters
of $M$ are tensely related to certain properties of the sequence
$M_1$, \ $M_2$, \ $M_3$, etc.

\subsection{ Family  of matrices invariant with
respect to the monodromization}  Denote by $\tau_{m,l}$, \
$m,l\in\N$, the set of the trigonometric polynomials of the form
$$f\,(z)=e^{\dsize -imz} f_{-m}+e^{\dsize -i(m-1)z}
f_{-m+1}+\dots+ e^{\dsize i(l-1)z} f_{l-1}+e^{\dsize ilz} f_{l},$$
e.i. such that $n_-(f)\le l$, \ $n_+(f)\le m$.

\smallpagebreak Fix $n\in \N$. By $\Omega\,(n)$ or, briefly, by
$\Omega$, denote the set of matrix functions $M\,(z)\in
SL\,(2,\C)$ such that $$\begin{array}{ll}a\in \tau_{n,n},\quad &
b\in \tau_{n,n-1},\\
      c\in \tau_{n-1,n},\quad &d\in \tau_{n-1,n-1},\end{array}$$
and
$$n_+(a)=n_-(a)=n.$$

\smallpagebreak Note that the conditions \eqref{1.3} and
\eqref{1.20} of the existence theorem are fulfilled for any matrix
from the family $\Omega$.

\smallpagebreak
One has

\medpagebreak {\bf Theorem 1.6.} {\it Let  $M\in \Omega\,(n)$.
Consider two canonical bases $f_{1,2}$ and $g_{1,2}$. Assume that
the corresponding minimal entire solutions $\psi_D$ and $\psi_B$
exist and that their wronskian is nonzero. Then the corresponding
monodromy matrix as a function of the variable $ z_1=2\pi z/h$
also belongs to $\Omega\,(n)$. }

\medpagebreak This theorem can be used as the base for a program
for the spectral investigation of the difference Schr\"odiner
equations with the trigonometric polynomial potentials: the
equation
$$\frac{\psi\,(z+h)+\psi\,(z-h)}2+p\,(z)\,\psi\,(z)=E\psi\,(z)$$
is equivalent to \eqref{1.1} with the matrix
$\left(\begin{array}{cc} 2E-2p\,(z) & -1\\ 1 &
0\end{array}\right)$. In this case, \ $M\in\Omega\,(n)$, and the
monodromization procedure leads to a sequence of equations
\eqref{1.1} with the matrices from the family $\Omega\,(n)$. And
thus, it is equivalent to a finite dimensional dynamical system.
In this paper, we do not consider any spectral problems, and note
only that the monodromization  procedure is close to the
asymptotic renormalization approach suggested for \eqref{1.2} by
Helffer and Sj\"ostrand \cite{HS} and Wilkinson \cite{W} under
some semiclassical assumptions on the number $h$.

\medpagebreak We finish this discussion by listing the asymptotic
formulae for the coefficients of the monodromy matrix from Theorem
1.6:
\begin{eqnarray}
{\mathcal M}_{11}(z)=&\alpha_{2}^0\,e^{\dsize - 2\pi i\,n\,
z/h}\,(1+o\,(1)),&\quad z\to+i\infty, \label{1.24}\\
{\mathcal M}_{11}(z)=&\beta_{1}^0\,e^{\dsize + 2\pi i\,n\,
z/h}\,(1+o\,(1)),&\quad z\to-i\infty,\label{1.25}\\
{\mathcal M}_{12}(z)=&-\alpha_{2}^0\,e^{\dsize - 2\pi i\,n
\,z/h}\, \left(\frac{A_D}{A_B}+o\,(1)\right),&\quad
z\to+i\infty,\label{1.26}\\
{\mathcal M}_{12}(z)=&-\beta_{1}^0\,e^{\dsize + 2\pi i\,(n-1)\,
z/h}\, \left(\frac{D_D}{D_B}+o\,(1)\right),&\quad
z\to-i\infty,\label{1.27}\\
{\mathcal M}_{21}(z)=&\alpha_{2}^0\,e^{\dsize - 2\pi i\,(n-1)\,
z/h}\, \left(e^{4\pi^2i/h}\,\frac{B_B}{B_D}+o\,(1)\right),&\quad
z\to+i\infty,\label{1.28}\\
{\mathcal M}_{21}(z)=&\beta_{1}^0\,e^{\dsize + 2\pi i\,n\, z/h}\,
\left(\,\frac{C_B}{C_D}+o\,(1)\right),&\quad
z\to-i\infty,\label{1.29}\\
{\mathcal M}_{22}(z)=&-\alpha_2^0\,e^{\dsize  -2\pi i\,(n-1)\,
z/h}\,
\left(e^{4\pi^2i/h}\,\frac{A_D\,B_B}{B_D\,A_B}+o\,(1)\right),&\quad
z\to+i\infty,\label{1.30}\\
{\mathcal M}_{22}(z)=&-\beta_1^0\,e^{\dsize +2\pi i\,(n-1)\,
z/h}\, \left(\frac{D_D\,C_B}{D_B\,C_D}+o\,(1)\right), &\quad
z\to-i\infty.\label{1.31}
\end{eqnarray}
Here $\alpha_{1,2}^0$ and $\beta_{1,2}^0$ are the constants from
the asymptotic representations \eqref{1.16} -- \eqref{1.17} for
the Bloch multipliers of the solutions $f_{1,2}$ and $g_{1,2}$.
Note also that, since the wronskian of $\psi_D$ and $\psi_B$  is
nonzero, then, in view of Proposition 1.3,
$A_B,\,D_B,\,B_D,\,C_D\ne0$.
\subsection{ Typical properties of the minimal
solutions} Above, we have described the main results of this
paper. We shall continue it in the next publication. There, we
shall study some typical properties of the minimal solutions and,
in particular, of their asymptotic coefficients. In fact, the
second part will be devoted to the study of the dependence of  the
minimal solutions and their asymptotic coefficients on the
parameters of the matrix $M$. Let us formulate here the central
results of this second part.

\medpagebreak
Together with the set $\Omega\,(n)$, we consider also its subsets
$\Omega_{ml}$, \ $-n\le -m\le l \le n-1$,
singled out by the conditions
$$n_+(b)= m,\quad n_-(b)=l.$$

\smallpagebreak
Clearly,
$$\Omega=\bigcup_{m,l} \Omega_{ml},\quad\quad {\rm and}\quad\quad
\Omega_{ml}\cap \Omega_{m'l'}=\emptyset,\quad (m,l)\ne (m',l').$$

\smallpagebreak It appears that the elements of $\Omega$ can be
uniquely parametrized by the constant coefficients of the
trigonometric polynomials $a\,(z)$ and $b\,(z)$. Denote by
$\omega$ the set of constant coefficients of the matrix $M$. One
can consider $\omega$ as a point in $\C^{8n}$. With this, the set
$\Omega_{ml}$, \ $-n\le -m\le l \le n-1$, appears to be a
connected analytic submanifold of $\C^{8n}$ of the dimension
$2n+m+l+2$, and the set $\Omega\,(n)$ itself is a connected
analytic submanifold of $\C^{8n}$ of the dimension $4n+1$.

\medpagebreak Let ${\mathcal D}$ be  a simply  connected domain of
$\Omega_{ml}$. Fix some continuous on ${\mathcal D}$ branches of
the functions $$\phi_-=-l\,h/2-i\ln a_{+n},\quad
\phi_+=-m\,h/2+i\ln a_{-n}.$$ Assume that, for any point
$\omega\in{\mathcal D}$, these functions satisfy the condition
\eqref{1.18}, e.i. that the canonical bases with these parameters
are consistent for all $\omega\in{\mathcal D}$. One has

\medpagebreak {\bf Theorem 1.7.} \ {\it For the above set
$\mathcal D$ and the above canonical bases, the minimal entire
solutions can be normalized so that they would be meromorphic in
$\omega\in{\mathcal D}$ together with their asymptotic
coefficients. In this case, the asymptotic coefficients are not
identically zero.}

\medpagebreak This theorem and formulae \eqref{1.26} --
\eqref{1.27} immediately imply

\medpagebreak {\bf Corollary 1.8.} \ {\it In the case of Theorem
1.7, the monodromy matrix corresponding to the minimal solutions
$\psi_D$ and $\psi_B$ is meromorphic in $\omega\in{\mathcal D}$.
Moreover, this matrix typically belongs to $\Omega_{n,n-1}$.}

\medpagebreak
Here, we have used the usual terminology of the analytic set theory:
one says that a property of a function defined on an analytic set is
{\it typical} if it takes place outside some analytic subset of a
smaller dimension.
\subsection{ The plan of the paper}  In section 2,
we prove Theorem 1.1. Also, we study some additional properties of
the Bloch solutions.

Section 3 is devoted to the scalar equation
\begin{equation} \mu\,(z+h)+\mu\,(z-h)+2\lambda\cos z\,\, \mu\,(z)=0,\quad z\in
\C,\label{1.32}\end{equation} where $h>0$ and $\lambda\in\C$ are
two fixed parameters. We construct and study minimal entire
solutions of this equation. The solutions are explicitly described
in terms of certain contour integrals. Note that the case of
$\lambda=-1$ was already treated in \cite{BF6}.

Section 4 is devoted to the analysis of the equation
\begin{equation} f\,(z+h)+f\,(z-h)+2\lambda\cos z f\,(z)=w\,(z)\,f\,(z),\quad
z\in \C, \label{1.33}\end{equation} where $w$ is a meromorphic
function satisfying the estimates $$|w\,(z)|\le
\Const\,e^{-(1-\epsilon)\,|\im z|},\quad \epsilon>0,$$ for
sufficiently big $|\im z|$. In terms of the minimal solutions of
\eqref{1.32}, we invert the operator in the left-hand side of
\eqref{1.33}, and obtain an integral Fredholm equation on a
contour in the complex plane. This allows to construct and
investigate some meromorphic solutions of \eqref{1.33}.

In Section 5, we  reduce the matrix equation \eqref{1.1} to the
form \eqref{1.33}, and then prove Theorem 1.2.

In the first part of Section 6, we study basic properties of the
minimal entire  solutions assuming that their asymptotic
coefficients are nonzero. In particular, we prove their linear
independence over the field of $h$-periodic function, their
uniqueness (up to independent of $z$ factors), and justify Theorem
1.6. Recall that the monodromy matrices corresponding to the
minimal solutions $\psi_D$ and $\psi_B$ are trigonometric
polynomials of $2\pi z/h$. Formulae \eqref{1.24} -- \eqref{1.31}
allow us to calculate only some of the constant coefficients of
these polynomials. In the end of the section, we describe a way to
calculate all the other coefficients.

Section 7 is devoted to some remarks for the case where $M$ is a trigonometric
polynomial of the first order.
\subsection{Acknowledgments} The work was accomplished when the
authors enjoyed the hospitality of the Universities of Paris XII
Val de Marne and Paris Nord. The authors are very grateful to the
professors Alain Grigis and Colette Guillop\'e for their aid. The
work was partially supported by the grant INTAS-RFBR 96-0414. The
results of the present paper were announced in \cite{BF8}, see
also \cite{BF9}.
\section{ Bloch solutions in a vicinity of the
infinity} In this section, we construct Bloch solutions $f_{1,2}$
of equation \eqref{1.1} described in Theorem 1.1a. Subsections 2.1
-- 2.5 are devoted  to the proof of this theorem.

In subsection 2.6, we discuss the set of all the solutions
analytic in $\C_+$ \ ($\C_-$) and having there the same asymptotic
representations as $f_{1,2}$ \ ($g_{1,2}$).
\subsection{ The plan of the proof}
Consider equation \eqref{1.5} for the first component $\psi_1$ of
a vector solution $\psi$ of equation \eqref{1.1}. Setting
\begin{equation} \Phi\,(z)=\psi_1(z+h)/\psi_1(z),\label{2.1}\end{equation} we see
that \begin{equation}
\Phi\,(z)+\rho\,(z)/\Phi\,(z-h)=v(z).\label{2.2}
\end{equation}
We call \eqref{2.2} a difference Ricatti equation.

To construct a Bloch solution of \eqref{1.1} analytic in a
vicinity of $+i\infty$ we use the ideas of \cite{BF5}. The first
step is to prove the existence of an analytic $2\pi$-periodic
solution $\Phi$ of the difference Ricatti equation in a vicinity
of $+i\infty$. Then we introduce the function
\begin{equation} \phi\,(z)=\ln\Phi\,(z)\label{2.3}\end{equation} and consider the
equation
\begin{equation} \lambda\,(z+h)-\lambda\,(z)=\phi\,(z).\label{2.4}\end{equation}
Having solved this equation we reconstruct the first component of
a vector solution of \eqref{1.1} by the formula \begin{equation}
\psi_1(z)=e^{\dsize \lambda\,(z)},\label{2.5}\end{equation} after
that the second one can be recovered by \eqref{1.6}. Finally, we
check that the $2\pi$-periodicity of $\Phi$ implies that the
constructed solution is a Bloch solution.
\subsection{ Analytic solutions of the Ricatti equation}
\medpagebreak {\bf Proposition 2.1.} \  {\it In the case of
Theorem 1.1a, there exist two $2\pi$-periodic solutions
$\Phi_{1,2}$ of \eqref{2.2} analytic in a vicinity of $+i\infty$
and having the asymptotics:
\begin{equation} \Phi_{2}=v\,(z)+o\,(1),\quad z\to
+i\infty,\label{2.6}\end{equation} and
\begin{equation} \Phi_{1}(z)=\frac{\rho\,(z+h)}{v\,(z+h)}\,(1+o\,(1)),\quad
z\to+i\infty.\label{2.7}\end{equation} }

\demo Consider the sequence of the functions
$$\Phi^{(0)}(z)=v\,(z),\quad
\Phi^{(1)}(z)=v\,(z)-\frac{\rho\,(z)}{v\,(z-h)},$$
$$\Phi^{(n)}(z)=v\,(z)-\frac{\rho\,(z)}{\Phi^{(n-1)}(z-h)}\ ,\quad
n\ge1. $$ The limit of this sequence (if exists) is a solution of
\eqref{2.2} represented by the continuous fraction
\begin{equation} \Phi_{2}(z)=
             v\,(z) -\frac{\dsize    \rho\,(z)}   {\dsize
             v\,(z-h) -\frac{\dsize  \rho\,(z-h)} {\dsize
             v\,(z-2h) -\frac{\dsize \rho\,(z-2h)}{\dsize
             \quad\cdots}}} \,.
\label{2.8}\end{equation} This continuous fraction was
investigated in \cite{BF5}.

\smallpagebreak Recall that the coefficients of the matrix $M$ are
trigonometric polynomials, and that $b\not\equiv0$. Both the
functions $v$ and $\rho$ are analytic in a vicinity of $+i\infty$.
Moreover, for $z\to+i\infty$, the function $\rho$ tends to a
finite limit, and $v\,(z)\to \infty$. Let $$v_-(Y)=\inf_{\im
z>Y}|v\,(z)|,\quad \rho_+(Y)=\sup_{\im z>Y}|\rho\,(z)|,\quad
\mu=\frac{\rho_+}{(v_-/2)^2}.$$ \smallpagebreak As in \cite{BF5},
one can easily check that if $\mu<1$, then
$$|\Phi^{(n+1)}(z)-\Phi^{(n)}(z)|\le \mu\,
|\Phi^{(n)}(z)-\Phi^{(n-1)}(z)|,\quad \im z>Y.$$ This estimate
implies that all the functions $\Phi^{(n)}$, \ $n=1,\,2,\,\dots$,
are analytic  near the "point" $+i\infty$, that the continuous
fraction in \eqref{2.8} converges uniformly in $z$ in the
half-plane $\C_+(Y)$ for sufficiently big $Y$, and that $\Phi_{2}$
satisfies the estimate  \eqref{2.6}. Since any of the functions
$\Phi^{(n)}$ is $2\pi$-periodic, the limit is also
$2\pi$-periodic.

The solution $\Phi_{1}$ can be represented by the continuous fraction
$$ \Phi_{1}(z)=
             \frac{\dsize    \rho\,(z+h)}   {\dsize
             v\,(z+h) -\frac{\dsize  \rho\,(z+2h)} {\dsize
             v\,(z+2h) -\frac{\dsize \rho\,(z+3h)}{\dsize
             \quad\cdots}}} \,.
$$ It can be investigated in the same way as \eqref{2.8}. \qed
\subsection{ Logarithms of the solutions of the Ricatti equation}
 Let
\begin{equation} \phi_{1,2}(z)=\ln\Phi_{1,2}(z),\label{2.9}\end{equation} where
$\Phi_{1,2}$ are the solutions of the Ricatti equation constructed
in Proposition 2.1. Asymptotic representations \eqref{2.6} --
\eqref{2.7} immediately imply

\medpagebreak {\bf Lemma 2.2.} \ {\it In a vicinity of $+i\infty$,
the functions $\phi_{1,2}(z)$ are analytic and  can be represented
in the form \begin{equation} \phi_{1,2}(z)=\pm in_+(v)\,z
+\alpha_{1,2} +g_{1,2}(z),\label{2.10}\end{equation} where
\begin{equation} \alpha_{2}=\ln\,v_+,\quad \alpha_{1}=-\ln\,v_+
-in_+(b)\,h+in_+(v)\,h,\label{2.11}\end{equation} and $g_{1,2}$
are $2\pi$-periodic analytic functions decaying as
$z\to+i\infty$.}
\subsection{ Homological equation}
Here, we collect some facts on the homological equation
\begin{equation} \varphi\,(z+h)-\varphi\,(z)=g\,(z),\quad
z\in\C,\label{2.12}\end{equation} where $g$ is a $2\pi$-periodic
function.

\medpagebreak
{\bf 1.} \ First, prove

\medpagebreak {\bf Lemma 2.3.} \ {\it Let $g$ be a $2\pi$-periodic
function analytic in a vicinity of $+i\infty$ decaying as
$z\to+i\infty$, then equation \eqref{2.12} has a solution analytic
in a vicinity of $+i\infty$ and decaying as $\im z\to +\infty$
uniformly in $\re z$ if $|\re z|$ is bounded by a constant}.

\demo The solution $\varphi$ can be constructed by explicit formulas.
Fix a point $z_o$ in the domain of analyticity of the function $g$. Let
$$G\,(k)=\frac1{2\pi}\,\int_{z_o}^{z_o+2\pi}e^{\dsize-ikz}\,g\,(z)\,dz,
$$
and
\begin{equation} \varphi\,(z)=
\frac1{2i}\,\int_{\gamma}G\,(k)\,
                    e^{\tsize ikz}
                    \frac{\ctg\pi k}{e^{\tsize ikh}-1}\,dk,\label{2.13}
                    \end{equation}
Here $\gamma$ is a contour in the complex plane of $k$. Describe it.

\smallpagebreak The integrand in \eqref{2.13} has poles only at
the real line. One of the poles is at $k=0$. One is situated at
the point $k_o=\min\{1,\,2\pi/h\}$. All the other are located
outside the interval $[0,k_0]$. Fix two constants $c_1$ and $c_2$
so that $c_1>0$, \ $0<c_2<k_o$. The contour $\gamma$ is  a contour
coming from $+\infty$ along the line $\im z=c_1$, going from the
upper half-plane of the complex plane along the line $\re z=c_2$,
and coming back to $+\infty$ along the line $\im z=-c_1$.

\smallpagebreak If $\im z>\im z_o$, the integral for $\varphi$
converges absolutely. Moreover, $$|\varphi\,(z)|\le
\Const\,e^{\dsize -c_2\,\im z}.$$ This estimate is uniform in $\re
z$ if it is bounded by a fixed constant.

\smallpagebreak One checks that $\varphi$ satisfies equation
\eqref{2.12} by means of the residue theorem:
$$\varphi\,(z+h)-\varphi\,(z)= \frac1{2i}\int_\gamma
G\,(k)\,e^{ikz}\ctg(\pi k)\,dk= \sum_{\kappa=1}^\infty e^{i\kappa
z}\,G\,(\kappa)=g\,(z).$$ \qed

\medpagebreak {\bf 2.} \ Note that if $\varphi$ is a solution of
\eqref{2.12}, then the function $\tilde
g\,(z)=\varphi\,(z+2\pi)-\varphi\,(z)$ is $h$-periodic:
\begin{eqnarray*}
\tilde g\,(z+h)-\tilde g\,(z)&=&\varphi\,(z+2\pi+h)-\varphi\,(z+h)-
\varphi\,(z+2\pi)+\varphi\,(z)=\\
 &=&g\,(z+2\pi)-g\,(z)=\\
 &=&0.
\end{eqnarray*}
\subsection{ Bloch solutions in a vicinity of $+i\infty$}
Now, we have all the ingredients to construct two Bloch solutions
of \eqref{1.1} analytic in a vicinity of the point $+i\infty$.
\subsubsection{First solution}   Let $\Phi_{2}$ be
the solution of the Ricatti equation \eqref{2.2}  described by
Proposition 2.1 and let $\phi_{2}$ be the logarithm of $\Phi_{2}$.
Consider the equation
\begin{equation} \lambda\,(z+h)-\lambda\,(z)=\phi_{2}(z).\label{2.14}\end{equation}
Remind that $\phi_{2}$ admits the representation \eqref{2.10}.
Therefore, any solution of \eqref{2.14} can be represented in the
form
\begin{equation} \lambda(z)= -in_+(v)\,\frac{z^2}{2h}+in_+(v)\frac z2+\ln
v_+\,\frac zh+ \varphi(z),\label{2.15}\end{equation} where
$\varphi$ is a solution of the homological equation  with
$g_{2}(z)$ in the right-hand side. We construct a solution
$\varphi$ of this equation as in Lemma 2.3, and then, we construct
a solution $\psi\,(z)$ of \eqref{1.1} by formulae \eqref{2.5} and
\eqref{1.6}. The result can be written in the form
\begin{equation} \psi\,(z)=e^{\lambda\,(z)}\,\left(\begin{array}{c} 1 \\
\frac{\Phi_2(z)-a\,(z)}{b\,(z)}\end{array}
\right).\label{2.16}\end{equation}

\smallpagebreak Note that since $b\,(z)$ is a trigonometric
polynomial, the function $\frac1{b\,(z)}$ is analytic in a
vicinity of $+i\infty$, and, therefore, $\psi$ is also analytic in
some vicinity $\C_+$ of $+i\infty$.

\smallpagebreak Check that $\psi$ is a Bloch solution. Since the
vector in the right hand side of \eqref{2.16} is $2\pi$-periodic,
$$\psi(z+2\pi)=\gamma\,(z)\,\psi\,(z),\quad z\in \C_+,$$ where
$$\gamma\,(z)=e^{\dsize \lambda(z+2\pi)-\lambda(z)}.$$ Thus, it
suffices to show that
\begin{equation} \gamma\,(z+h)=\gamma\,(z).\label{2.17}\end{equation}

\smallpagebreak By \eqref{2.15}, $$\gamma\,(z)= {\rm
Const}\,e^{\dsize -n_+(v)\,\frac{2\pi i z}h} e^{\dsize
\varphi(z+2\pi)-\varphi(z)}.$$ But, in view of section 2.3.2, the
function $\varphi(z+2\pi)-\varphi(z)$ is $h$-periodic. This
implies the $h$-periodicity of $\gamma$.

Denote the constructed solution by $f_2$.
We have come to

\medpagebreak {\bf Lemma 2.4.} \ {\it There exists a Bloch
solution $f_2$ of \eqref{1.1} analytic in a vicinity of $+i\infty$
with the first component admitting the representation:
$${(f_2)}_1(z)=e^{\dsize -in_+(v)\,\frac{z^2}{2h}+\ln v_+\,\frac
zh+in_+(v)\frac z2+o\,(1)},\quad \im z\to+\infty.$$ }
Note that, here, $o(1)$ is just the function $\varphi$ decaying as
$\im z\to +i\infty$.
\subsubsection{Second solution}  Starting with
$\Phi_{1}$ and repeating the above arguments we  obtain

\medpagebreak {\bf Lemma 2.5.} \ {\it There exists a Bloch
solution $f_1$ of \eqref{1.1} analytic in a vicinity of $+i\infty$
with the first component admitting the representation:
$${(f_1)}_1(z)=e^{\dsize in_+(v)\,\frac{z^2}{2h}-\ln v_+\,\frac
zh+ in_+(v)\,z/2-in_+(b)z+o\,(1)},\quad \im z\to+\infty.$$ }
\subsubsection{ Linear independence}  Calculate the
determinant of the matrix $F\,(z)\equiv$ $
\left(f_1(z),f_{2}(z)\right)$. Take advantage of formula
\eqref{1.8}. This gives $$\det
F\,(z)=\frac{{(f_1)}_1(z){(f_2)}_1(z)}{b\,(z)}\,
                \left(\Phi_{2}(z)-\Phi_{1}(z)\right).$$
Substituting in this formula the asymptotic representations for
$f_{1,2}$ and $\Phi_{1,2}$, we come to the representation
\begin{equation} \det F\,(z)= v_+/b_+(1+g\,(z)),\quad\quad
g\,(z)\to 0,\quad z\to+i\infty. \label{2.18}\end{equation} Here
$g$ is an $h$-periodic function, analytic in a vicinity of
$+i\infty$. Note that formula \eqref{2.18} implies that $f_1$ and
$f_2$ are linearly independent over the field of $h$-periodic
functions in a vicinity of $+i\infty$.

\bigpagebreak Now,  we redefine the Bloch solutions $f_1$ and
$f_2$: $$f_1:=q\,f_1/(1+g), \quad f_2:=f_2/q,\quad
q=\exp(i\phi_+^2\,/\,(2hn_+(v))).$$ The new $f_1$ and $f_2$ are
also Bloch solutions of equation \eqref{1.1}, but their wronskian
is constant. They are the solutions  described in Theorem 1.1a.
The proof of Theorem 1.1b is absolutely similar. \qed
\subsection {Uniqueness of the Bloch solutions}
\ {\bf 1.} \ First, we answer the question:" Are the Bloch
solutions described in Theorem 1.1 uniquely defined  by  their
asymptotics for $z\to\pm i\infty$ ?"

\bigpagebreak {\bf Lemma 2.6.} \ {\it Let $\tilde f$ be a Bloch
solution of \eqref{1.1} analytic in some vicinity $\C_+$ of
$+i\infty$ and, as the solution $f_1$, having the asymptotics
\eqref{1.10} for $z\to+i\infty$. Then $$\tilde
f(z)=c\,(z)\,f_1(z),$$ where $c\,(z)$ is an $h$-periodic function
analytic in a vicinity of $+i\infty$, and having the asymptotics
$$c\,(z)\to1,\quad z\to+i\infty.$$ } \demo Since the solutions
$f_{1,2}$ are linearly independent in some vicinity of $+i\infty$,
\ $\tilde f$ can be represented by their linear combination with
some $h$-periodic analytic coefficients $$\tilde
f(z)=\alpha\,(z)\,f_1(z)+\beta\,(z)\,f_2(z).$$ The coefficients
$\alpha$ and $\beta$ can be expressed explicitly in terms of the
solutions: $$\alpha=\frac{\det\,(\tilde
f,f_2)}{\det\,(f_1,f_2)},\quad
  \beta=\frac{\det\,(f_1,\tilde f)}{\det\,(f_1,f_2)}.$$

\smallpagebreak Since both $f_1$ and $\tilde f$ are described by
\eqref{1.10} as $z\to+i\infty$, $$\alpha\,(z)\to 1,\quad
z\to+i\infty.$$

\smallpagebreak Now, show that $\beta\equiv 0$. Since $\tilde f$
and $f_1$ are Bloch solutions $$\det\,(f_1,\tilde f)\,(z+2\pi N)=
\det\,(f_1,\tilde f)\,(z)\, \prod_{l=0}^{N-1}u_1(z+2\pi l)\,\tilde
u\,(z+2\pi l), \quad \forall\,\,N\in\N,$$ where $u_1$ and $\tilde
u$ are the Bloch coefficients of $f_1$ and $\tilde f$. The
asymptotics \eqref{1.10} (see also Remark 2 to Theorems1.1) imply
that, for sufficiently big $\im z$, $$|u_1(z)|,\,|\tilde
u\,(z)|\le Ce^{-2\pi n_+(v) \im z\,/\,h}.$$ This formula is
uniform in $\re z$ since \eqref{1.10} is uniform in $\re z$ if
$|\re z|$ is bounded by a constant, and since $u_1$ and $\tilde u
$ are $h$-periodic. In result, for sufficiently big $\im z$,
$$|\det\,(f_1,\tilde f)\,(z+2\pi N)|\le C\,e^{-4\pi N n_+(v)\im
z/h} |\det\,(f_1,\tilde f)\,(z)|,\quad \forall\,\,N\in\N, $$
uniformly in $\re z$. Since the determinant is $h$-periodic, this
is possible only if it equals to $0$. Thus, $\tilde
f(z)=\alpha\,(z)\,f_1(z)$ which proves the lemma. \qed

\bigpagebreak
We have investigated the uniqueness of the solution $f_1$.
For the solutions $f_2$, $g_1$ and $g_2$ one can prove the similar statements.

\bigpagebreak {\bf 2.} \ In constructing the Bloch solutions, we
have fixed the parameters $\phi_\pm$ defined by formulae
\eqref{1.11} and \eqref{1.14} only modulo $2\pi$. The choice of
the parameter $\phi_+$ \ ($\phi_-$) influences the asymptotics of
$f_{1,2}$ \ ($g_{1,2}$) as $z\to+i\infty$ \ ($z\to -i\infty$).
Fixing this parameter in two different ways, we obtain two
different bases $f_{1,2}$ \ ($g_{1,2}$). Let us study relations
between these bases. We shall indicate explicitly the dependence
of $f_{1,2}$ on $\phi_+$. On has

\bigpagebreak {\bf Lemma 2.7.} \  {\it \begin{equation}
f_{1,2}(z,\phi_+ + 2\pi)=c_{1,2} e^{\dsize \pm2\pi i
z/h}\,f_{1,2}(z,\phi_+), \label{2.19}\end{equation} where
$c_{1,2}=c_{1,2}(z,\,\phi_+)$ are some analytic $h$-periodic
functions tending to nonzero constants as $z\to+i\infty$.}

\medpagebreak
The proof of this lemma is similar to the proof of the previous one.
Again, one can prove the same statement for the Bloch solutions
$g_{1,2}$.

\smallpagebreak
The most interesting case is one where we change $\phi_+$ by $2\pi n_+(v)$.
In this case, one comes to

\medpagebreak {\bf Corollary 2.8.} \ {\it \begin{equation}
f_{1,2}(z,\phi_+ + 2\pi n_+(v))=C_{1,2} f_{1,2}(z+2\pi\,,\phi_+),
\label{2.20}\end{equation} where $C_{1,2}=C_{1,2}(z,\,\phi_+)$ are
some analytic $h$-periodic functions with the asymptotics
$$C_{1,2}(z,\,\phi_+)=e^{\dsize-\pi i(n_+(v)-n_+(b))}+o\,(1),
\quad z\to+i\infty.$$}
\section{ Minimal solution of the auxiliary equation}
The next three sections are devoted to constructing of the minimal
solutions of equation \eqref{1.1}. Let us agree on the terminology
and the notations we shall use there.

For $z \in\C$, we call the set $$\{\zeta\in\C\,: \,\im z=\im
\zeta,\,\, -\delta<\re(z-\zeta)<\delta\}$$ {\it the horizontal
$\delta$-vicinity of $z$}. We call {\it the horizontal
$\delta$-vicinity of a curve} $\gamma\subset\C$ the union of the
horizontal $\delta$-vicinities of the all its points. We call {\it
the horizontal distance} between a curve $\gamma$ and a point $z$
the maximal  $\delta$ for which the horizontal $\delta$-vicinity
of $\gamma$ does not contain $z$.

We call a curve $\gamma\subset \C$ {\it vertical}, if it
intersects all the lines $\im z={\rm Const}$ only at nonzero
angles. If all these angles are strictly bigger than some fixed
positive constant, we call the curve {\it strictly vertical}.

Furthermore, in the sequel, we use the letter $C$ as a symbol denoting
constants independent of $z$.

\medpagebreak In  Section 3, we construct and investigate minimal
entire solutions of the model equation
\begin{equation}\label{3.0}
\left(\begin{array} {c}m_1(z+h)\\ m_2(z+h)\end{array}\right)=
\left(\begin{array}{cc} -2e^\xi \cos z & -1\\1 &
0\end{array}\right)\,\left(\begin{array}{c} m_1(z)\\
m_2(z)\end{array}\right),\quad z\in\C,
\end{equation}
where $\xi$ is a complex number. Note that the first component
$m_1$ of its vector solutions satisfies the equation
\begin{equation} m\,(z+h)+m\,(z-h)+2e^{\xi}\,\cos
z\,m(z)=0,\label{3.1}\end{equation}
and the second component is related to $m_1$ by the formula
$m_2(z)=m_1(z-h)$. The main results are formulated in Theorem 3.2.
When describing the results, we use the special function described
in Section 3.2.

In \cite{BF6}, we have considered the case where  $\xi=0$. Here,
we discuss in details only the new elements of the proof.
\subsection{ Reduction to a first order difference
equation}\label{S:3.1}
 We shall construct a solution $m(z)$ of \eqref{3.1}
in the form:
\begin{equation} m\,(z)=e^{\dsize -iz^2/2h}\,\int_{\Gamma}
e^{\dsize -izp/h -ip^2/4h+\pi i p/2h}\,
v\,(p)\,dp,\label{3.2}\end{equation}
where $\Gamma$ is a contour in the complex plane of $p$.
Substituting \eqref{3.2} into \eqref{3.1}, we formally get
\begin{eqnarray*}
\lefteqn{e^{-iz}\int_{\Gamma} e^{\dsize -izp/h -ip^2/4h+\pi i
p/2h}\,\left(e^{-ip-ih/2}+e^\xi\right) v\,(p)\,dp+}   \ \ \ \\ &
&\ \ \ \  e^{+iz}\int_{\Gamma} e^{\dsize -izp/h -ip^2/4h+\pi i
p/2h}\,\left(e^{ip-ih/2}+e^\xi\right) v\,(p)\,dp=0,
\end{eqnarray*}
or, after having changed the variables,
\begin{eqnarray*}
\lefteqn{ \int_{\Gamma+h} e^{\dsize -izp/h -i(p-h)^2/4h+\pi i
(p-h)/2h}\,\left(e^{-ip+ih/2}+e^\xi\right) v\,(p-h)\,dp+}   \ \ \
\\ & & \ \ \ \int_{\Gamma-h} e^{\dsize -izp/h -i(p+h)^2/4h+\pi i
(p+h)/2h}\,\left(e^{ip+ih/2}+e^\xi\right) v\,(p+h)\,dp=0,
\end{eqnarray*}
where $\Gamma\pm h$ are the contours obtained of $\Gamma$ by $\pm
h$-translations. Assuming that $v$ is analytic in a sufficiently
large vicinity of $\Gamma$, and that there are no problems of
convergence of the integrals, we  deform the integration contours
back to $\Gamma$ and obtain
$$\int_{\Gamma} e^{ -izp/h -ip^2/4h+\pi i
p/2h}\left[\left(e^{-ip/2+ih/2}+e^{\xi+ip/2}\right) v(p-h)-
 \left(e^{ip/2+ih/2}+e^{\xi-ip/2}\right) v(p+h)\right]dp=0.$$
So, if $v$ satisfies the first order difference equation
\begin{equation}
v\,(p+h)=\varrho\,(p)\,v\,(p-h),\label{3.3}\end{equation} with
$$\varrho\,(p)=\frac{e^{\dsize ih/2}+e^{\dsize ip+\xi}} {e^{\dsize
\xi}+e^{\dsize ih/2+ip}},$$
then $m$ is a solution of \eqref{3.1}. Now, our aim is to
construct a solution of \eqref{3.3}. To get a detailed information
about this solution we shall need a special function playing an
important role in the analytic theory of difference equations with
periodic coefficients.
\subsection{ $\sigma$-function}
Here, we construct a meromorphic solution of the equation
\begin{equation} \sigma\,(z+h)=(1+e^{\dsize
-iz})\,\sigma\,(z-h),\quad z\in \C.\label{3.4}\end{equation} A
similar function was introduced and systematically used in the diffraction theory
\cite{BoFi}. Later, it was also introduced and studied in 
a different context by other authors, see, for example,  \cite{FKV} and \cite{R}.
So, we discuss this function omitting most of the proofs.

\smallpagebreak {\bf 1.} \ Let $\C_\pi$ be the complex plane cut
along the real line from $-\infty$ to $-\pi$, and from $\pi$ to
$+\infty$. Fix on $\C_\pi$ a branch of the function
$l_0\,(z)=\ln\,(1+e^{\dsize-iz})$ by the condition
\begin{equation}  l_0(z)\to 0,\quad z\to-i\infty.\label{3.5}
\end{equation}
Define $$L_0(z)=\int_{-i\infty}^{z}l_0(z')\,dz',\quad
z\in\C_\pi,$$ where the integration contour belongs to $\C_\pi$.
Let \begin{equation} \theta_0(z)=
\frac{\pi}{8ih^2}\int\limits_{\gamma} \frac {\tsize L_0(z')}
{\dsize \cos^2 \left(\frac
{\pi(z-z')}{2h}\right)}\,dz'.\label{3.6}\end{equation} Here,
$\gamma\subset\C_\pi$  is a strait line $\re z=\Const$ passing
between the points $z\pm h$. The function $\theta_0$ is analytic
in the strip $S_0=\{z\in\C:\,-\pi-h<\re z<\pi+h\}$.

\smallpagebreak By means of the residue theorem, one can easily
check that $$\theta_0(z+h)-\theta_0(z-h)=l_0(z), \quad z\pm h\in
S_0. $$ Therefore, the function $$\sigma\,(z)=e^{\dsize
\theta_0(z)}$$ is a solution of \eqref{3.4} in the strip $S_0$. By
means of \eqref{3.4}, one can continue it meromorphically in the
whole complex plane. Let now $\sigma$ denote this meromorphic
solution. Investigate its analytic properties.

\medpagebreak
{\bf 2.} \ Discuss the set of poles and zeros of the function $\sigma\,$.
By construction, the function $\sigma\,$ is analytic and has no zeros in
$S_0$. In result of the continuation, each  zero $z=\pi+2\pi l$,
$l=0,\,1,\,2,\,\dots$, of the function $1+e^{-iz}$  produces a chain of zeros
of $\sigma\,$ at the points
$$\pi+h +2\pi l+2hk,\quad k=0,\,1,\,2,\,\dots\,,$$
and each its zero $z=-\pi-2\pi l$, \ $l=0,\,1,\,2,\,\dots\,$,
produces a chain of poles
of $\sigma\,$ at the points
$$-\pi-h -2\pi l-2hk,\quad k=0,\,1,\,2,\,\dots.$$

\medpagebreak {\bf 3.} \ The pole $z=-\pi-h$ is simple. One can
calculate explicitly the residue of $\sigma$. Omitting long
elementary  calculations, we write down the result
\begin{equation} {\rm Res}_{z=-\pi-h} \sigma\,=-i\sigma\,(-\pi+h)=
\sqrt{\frac h\pi}\,e^{\dsize -\frac{i\pi^2}{12h}-\frac{i\pi}4
-\frac{ih}{12}}. \label{3.7}\end{equation} One can also get the
following explicit formula \begin{equation} \sigma\,(-\pi)=\frac
1{\sqrt{2}}\,e^{\dsize -i\pi^2\,/\,12h+ih\,/\,24}.
\label{3.8}\end{equation}

\medpagebreak {\bf 4.} \ Let us describe the asymptotics of
$\sigma\,$ for $\im z\to\pm i\infty$. Clearly,
$$L_0(z)=O\,(e^{\dsize -|\eta|}),\quad \im z\to-i\infty.$$ On the
other hand, $$L_0(z)=-iz^2/2+i\pi^2/6+O\,(e^{\dsize
-|\eta|}),\quad \im  z\to+i\infty.$$ This implies the asymptotics
\begin{equation} \sigma\,(z)=1+o\,(e^{-\mu|\im z|}),\quad \im
z\to-i\infty,\label{3.9}\end{equation} and \begin{equation}
\sigma\,(z)= e^{\dsize
-i\frac{z^2}{4h}+i\frac{\pi^2}{12h}+i\frac{h}{12}}\,
(1+o\,(e^{-\mu|\im z|})),\quad \im
z\to+i\infty,\label{3.10}\end{equation} where $\mu $ is a positive
number such that $\mu<{\rm min} \{1,\,\pi/h\}$. As before, we omit
elementary calculations, noticing only that the leading terms in
\eqref{3.10} can be obtained by the substitution in \eqref{3.6}
instead of the function $L_0$ the leading terms of its asymptotics
for $z\to+i\infty$.

\medpagebreak {\bf 5.} \ The solution $\sigma$ of equation
\eqref{3.4} is uniquely  determined by its asymptotics \eqref{3.9}
-- \eqref{3.10}, and by the fact that it is analytic and has no
zeros in the strip  $S_0$.

\smallpagebreak Indeed, let $\sigma_1$ be another solution of
\eqref{3.4} possessing these properties. By \eqref{3.4}, the ratio
$\sigma_1(z)/\sigma(z)$ a $2h$-periodic function. It is analytic
in $S_0$, and thus, due to the periodicity, it is entire. On the
other hand, it tends to $1$ as $z\to\pm i\infty$. Therefore, this
ratio is identically equal to $1$.

\medpagebreak {\bf 6.} \ The $\sigma$-function satisfies the
functional relations: \begin{equation}
\sigma\,(z+\pi)=(1+e^{\dsize -\frac{i\pi}h\,z})\,\sigma\,(z-\pi),
\label{3.11}\end{equation} \begin{equation} \sigma\,(-z)\,
=e^{\dsize -\frac{i}{4h}\,z^2+\frac{i\pi^2}{12}+
\frac{ih}{12}}\,\frac1{\sigma\,(z)},\label{3.12}\end{equation}
$$\overline{\sigma\,(\overline{z})}=\frac1{\sigma\,(- z)}.$$

\smallpagebreak These three relations can be proved by using
almost one and the same argument. To justify, for example, formula
\eqref{3.11}, first, one checks that the ratio
$r\,(z)=\sigma\,(z)/\sigma\,(z-2\pi)$  is a $2h$-periodic entire
function. Really, both the functions $\sigma(z)$ and
$\sigma(z-2\pi)$ satisfy \eqref{3.4}. This implies that $r$ is
$2h$-periodic. The function $\sigma(z)$ is analytic in the
half-plane $-\pi-h<\re z$, and $1/\sigma(z-2\pi)$ is analytic in
the half-plane $\re z<3\pi+h$. So, $r$ is analytic in the strip
$-\pi-h<\re z< 3\pi+h$. Thus, being $2h$-periodic, it is entire.
Furthermore, the asymptotics of $\sigma$ imply that $r\,(z)\to1$
as $z\to-i\infty$, and $r\,(z)=e^{\dsize -i\pi
z/h+i\pi^2/h}(1+o\,(1))$ as $z\to+i\infty$. This is possible only
if $r\,(z)=1+e^{\dsize \frac{i\pi^2}h-\frac{i\pi}h\,z}$, which
proves \eqref{3.11}.
\subsection{ Meromorphic solution of equation
\eqref{3.3}} Now, we come back to equation \eqref{3.3}. One of its
solution can be constructed by the formula \begin{equation}
v\,(p)=e^{\dsize
-i\frac{p_0}{2h}p}\,\sigma\,(p-p_0)/\sigma\,(p+p_0),
\label{3.13}\end{equation} where
\begin{equation} p_0=i\xi+h/2.\label{3.14}\end{equation} Describe analytic
properties of $v$.

\smallpagebreak First of all, note that \eqref{3.12} implies that
\begin{equation} v\,(-p)=v\,(p).\label{3.15}\end{equation}

\smallpagebreak The function $v$ has two chains of poles:
$$\pm(-p_o+\pi+h+2\pi j+2h k),\quad  j,k\in\N\cup\{0\}.$$ If these
two chains do not intersect, then the poles $\pm(-p_0+\pi+h)$ are
simple, and $$ \hskip-4cm
\Res_{p_0-\pi-h}v\,(p)=-\Res_{-p_0+\pi+h}v\,(p)=$$
\begin{equation} \hskip2cm =\sqrt{\frac h\pi}\,e^{\dsize
i\xi^2\,/\,2h
-\pi\xi\,/\,2h-i\pi^2\,/\,12h+ih\,/\,24}\,\frac1{\sigma(2i\xi-\pi)}.
\label{3.16}\end{equation}

\smallpagebreak The asymptotic formulae  for $\sigma$ imply
\begin{equation} v\,(p)=e^{\pm ip_0p\,/\,2h}(1+o\,(1)),\quad p\to\pm
i\infty.\label{3.17}\end{equation}
\subsection{ Minimal solution of the model equation
for $\re p_0<\pi$}
 In this section we assume that
\begin{equation}\label{3.181}\im \xi>-\pi+h/2. \end{equation}
 This condition is equivalent to
the inequality $\re p_0<\pi$. It implies that the poles of $v$ are
outside the strip
\begin{equation} -h\le\re p\le h.\label{3.18}\end{equation}

\noindent{\bf 1.} \ Construct an entire solution of the auxiliary
equation \eqref{3.1}. Let $\Gamma$ be a vertical curve going along
the imaginary axis from $p=-i\re \xi$ to $p=+i\re\xi$ and having
as the asymptotes for $p\to\pm i\infty$ the line $e^{\dsize
-i\pi/4}\,\R$. Since $v$ has the asymptotics \eqref{3.17}, the
integral in \eqref{3.2} converges and defines an entire function
$m\,(z)$.

\smallpagebreak This function $m$ is an entire solution of the
auxiliary equation \eqref{3.1}. To check this, we make the
calculation from subsection \ref{S:3.1}. As the convergence of the
integrals is obvious, then, to justify this calculation, one has
only to note that the horizontal distance from the contour
$\Gamma$ to the poles of $v$ is bigger than $h$.

\medpagebreak
{\bf 2.} \ Let us turn the attention to
the asymptotics of $m$ for $z\to\pm i\infty$. One has

\medpagebreak {\bf Proposition 3.1.} \ {\it Let $\im
\xi>-\pi+h/2$. Then the function $m$ has the asymptotics $$
\hskip-5cm m\,(z)= a_0e^{\dsize \frac
i{2h}\,(z-\pi+i\xi)^2+\frac{iz}2}\,(1+o\,(1))+$$ \begin{equation}
\hskip3cm +b_0e^{\dsize -\frac
i{2h}\,(z-\pi+i\xi)^2+\frac{iz}2}\,(1+o\,(1)), \quad z\to+i\infty,
\label{3.19}\end{equation}

$$ \hskip-5cm m\,(z)= c_0e^{\dsize \frac
i{2h}\,(z-\pi-i\xi)^2-\frac{iz}2}\,(1+o\,(1))+$$ \begin{equation}
\hskip2cm +e^{\dsize -2\pi iz/h}\, d_0e^{\dsize -\frac
i{2h}\,(z-\pi-i\xi)^2-\frac{iz}2}\,(1+o\,(1)), \quad
z\to-i\infty,\label{3.20}\end{equation} where $a_0$, $b_0$, $c_0$
and $d_0$ are independent of $z$. The asymptotics are uniform in
$\re z$ if $|\re z|$ is bounded by a fixed constant. The
asymptotic representations for $m'$ can be obtained by
differentiating the asymptotic representations for $m$.}

\medpagebreak When proving the proposition, one checks also that
the constant coefficients in the above asymptotics are given by
the formulae: \begin{equation} a_0=2i\sqrt{\pi h}\,e^{\dsize
-\frac i{4h}\,(i\xi-\pi)^2-
\frac14\xi+\frac{ih}{16}},\label{3.21}\end{equation}
\begin{equation} b_0= \frac{2\sqrt{\pi h}}{\sigma\,(2i\xi-\pi)}\,
e^{\dsize -\frac{i}{4h}\,(\pi-i\xi)^2 -\frac14\xi
-\frac{i\pi^2}{12h}-\frac{ih}{48}},\label{3.22}\end{equation}
\begin{equation} c_0= -2\sqrt{\pi h}e^{\dsize
-\frac{i}{4h}\,(\pi+i\xi)^2 -\frac14\xi+\frac{ih}{16}},
\label{3.23}\end{equation} \begin{equation}  d_0=\frac{2\sqrt{\pi
h}i}{\sigma\,(2i\xi-\pi)}\, e^{\dsize -\frac{2\pi\xi}h
-\frac{i}{4h}\,(\pi+i\xi)^2 -\frac14\xi+
\frac{11i\pi^2}{12h}-\frac{ih}{48}}. \label{3.24}\end{equation}

\medpagebreak For the case where $\xi\equiv 0$, we have proved
this proposition in \cite{BF6}. In the case under the
consideration, one can use the same arguments and, even, almost
the same estimates and calculations. So, we shall not repeat them
here, and give only some comments on the proof.

\medpagebreak
Begin with the case of $z\to+i\infty$.
Then, the asymptotics contains two leading terms. The first
one is defined by the behavior of $v$ as $p\to -i\infty$, and the second
one is related to the pole of the function $v$ situated at $p_0-\pi-h$.
Describe the first one.

\smallpagebreak Remind that as $p\to \pm i\infty$, the integrand
in \eqref{3.2} has the asymptotics $$e^{\dsize -\frac{iz}h\,p
-\frac i{4h}\,p^2+\frac{i\pi}{2h}\,p\pm\frac{ip_0}{2h}\,p}
(1+o\,(1)).$$ If $z\to+i\infty$, the integrand has a saddle point.
It satisfies the equation $$\dsize \frac{d}{dp}\left(
-\frac{iz}h\,p  -\frac
i{4h}\,p^2+\frac{i\pi}{2h}\,p-\frac{ip_0}{2h}\,p \right)=0,$$ and
thus, the saddle point is  $p_1=-2z+\pi-p_0$. The first of the
leading terms equals to the contribution of this saddle point to
\eqref{3.2}, e.i. to $$\hskip-4cm e^{\dsize
-\frac{i}{2h}\,z^2}\,\sqrt{4\pi h}\,e^{\frac{3\pi i}4}\, e^{\dsize
-\frac ih\,zp_1 -\frac i{4h}p_1^2+\frac{i\pi}{2h}\,p_1
-\frac{i}{2h}\,p_0p_1}=$$ $$\hskip+6cm =a_0e^{\dsize \frac
i{2h}\,(z-\pi+i\xi)^2+\frac{iz}2},$$ where $a_0$ is given by
\eqref{3.21}.

\smallpagebreak The second leading term of the asymptotics of
$m\,(z)$ equals to the contribution of the pole $p_0-\pi-h$ to
\eqref{3.2}, e.i. to $$\hskip-1cm e^{\dsize
-\frac{i}{2h}\,z^2}\,2\pi i\, \Res_{p=p_0-\pi-h} \left(e^{\dsize
-izp/h -ip^2/4h+\pi i p/2h}\, v\,(p)\right)=$$ $$\hskip+3cm
=b_0e^{\dsize -\frac i{2h}\,(z-\pi+i\xi)^2+\frac i2\,z},$$ with
$b_0$  from \eqref{3.22}.

\smallpagebreak The asymptotics \eqref{3.19} of $m\,(z)$ for
$z\to+i\infty$ can be described as the sum of the contributions of
the above two terms.

\smallpagebreak In the case where $z\to-i\infty$, the asymptotic
\eqref{3.20} of $m$ again contains two leading terms. The first
one is due to the saddle point $p_2$ defined by
$$\dsize \frac{d}{dp}\left( -\frac{iz}h\,p  -\frac
i{4h}\,p^2+\frac{i\pi}{2h}\,p+\frac{ip_0}{2h}\,p \right)=0,$$
i.e. $p_2=-2z+\pi+p_0$.  The contribution of this saddle point to
\eqref{3.2} is given by
$$\hskip-4cm e^{\dsize -\frac{i}{2h}\,z^2}\,\sqrt{4\pi h}\,
e^{\frac{3\pi i}4}\, e^{\dsize -\frac ih\,zp_2 -\frac
i{4h}p_2^2+\frac{i\pi}{2h}\,p_2 +\frac{i}{2h}\,p_0p_2}=$$
$$\hskip+6cm =c_0e^{\dsize \frac
i{2h}\,(z-\pi-i\xi)^2-\frac{iz}2},$$ where $c_0$ is given by
\eqref{3.23}. This is the first leading term, and the second one
equals to
$$\hskip-1cm  -e^{\dsize -\frac{i}{2h}\,z^2}\,2\pi i\,
\Res_{p=-p_0+\pi+h} \left(e^{\dsize -izp/h -ip^2/4h+\pi i p/2h}\,
v\,(p)\right)=$$ $$\hskip+3cm =e^{\dsize -2\pi iz/h}\,
d_0e^{\dsize -\frac i{2h}\,(z-\pi-i\xi)^2-\frac{iz}2}.$$

\qed

\medpagebreak {\bf 3.} \ Recall that we can reconstruct the
components of a vector solution of \eqref{3.0} by the formule
$m_1(z)=m(z)$ and $m_2(z)=m(z-h)$. This vector solution is a
minimal entire solution. Let us discuss this in detail.

\smallpagebreak The theorems 1.1a and 1.1b imply existence of the
canonical vector solutions $f_{1,2}$ and $g_{1,2}$ of \eqref{3.0}.
The asymptotics of their first components have the form
\begin{equation} (f_{1,2})_1(z)=e^{\dsize \pm\frac{i}{2h}\,(z+\phi_+)^2+
i\frac{z}2+o\,(1)},\quad z\to+i\infty,\label{3.26}\end{equation}
$$ \phi_+=\dsize i\xi -\pi,$$ and \begin{equation}
(g_{1,2})_1(z)=e^{\dsize \pm\frac{i}{2h}\,(z+\phi_-)^2-
i\frac{z}2+o\,(1)},\quad z\to-i\infty,\label{3.27}\end{equation}
$$\phi_-=-i\xi-\pi.$$
Here, we have chosen some of the possible values of the parameters
$\phi_\pm$. Comparing the asymptotics of $m_1(z)=m\,(z)$
(described by Proposition 3.1) with \eqref{3.26} -- \eqref{3.27} ,
we see that the vector $\left(\begin{array}{c}m_1(z)\\
m_2(z)\end{array}\right)$ is really a minimal solution of
\eqref{3.0} corresponding to the chosen canonical bases.
\smallpagebreak Below, we shall refer to $m$ as to a minimal
entire solution of \eqref{3.1}.
\subsection{ Minimal solution in the general case}
In the previous section, we have constructed the minimal entire
solution $m(z)$ for $\xi$ being in the half-plane \eqref{3.181}.
It is analytic in $\xi$ in this half-plane. Now, we are going to
continue it meromorphically on the whole complex plane of $\xi$
and, in result, to prove

\medpagebreak {\bf Theorem 3.2.} \ {\it Equation \eqref{3.1} has a
solution $m\,(z,\xi)$ entire in $z$ and meromorphic in $\xi$. It
has the following properties:

\point{i} $m$ is analytic in $\xi$ in the upper half-plane of the
complex plane $\xi$, and the function
$\sigma\,(2i\xi-\pi)\,m\,(z,\,\xi)$ is analytic in the half-plane
$\im\xi\le0$.

\point{ii} If $\sigma\,(2i\xi-\pi)\ne0$, then the behavior of $m$
for $z\to\pm i\infty$ is described by formulae \eqref{3.19} --
\eqref{3.20} with coefficients $a_0$, $b_0$, $c_0$ and $d_0$ given
by \eqref{3.21} -- \eqref{3.24}. }

\medpagebreak Above, we have indicated explicitly the dependence
of $m$ on $\xi$. The rest of the section is devoted to the proof.

\medpagebreak {\bf 1.} \ Let us study analytic properties of $m$
as a function of $\xi$. As we have already noted, it is analytic
in $\xi$ if $\im \xi>-\pi+h/2$. Continue meromorphically
$m\,(z,\,\xi)$ into the whole complex plane of $\xi$. The idea is
to find a relation connecting the functions $m\,(z,\,\xi)$,
$m\,(z,\,\xi+i\pi)$ and $m\,(z,\,\xi+2i\pi)$ for $\im
\xi>-\pi+h/2$, and to use this relation for the continuation. Such
a relation exists since all the three functions $m\,(z,\,\xi)$,
$e^{i\pi z/h}\,m\,(z,\,\xi+i\pi)$ and $m\,(z,\,\xi+2i\pi) $
satisfy \eqref{3.1}, and since the space of solutions of the
equivalent to \eqref{3.1} equation \eqref{3.0} is a two
dimensional module over the ring of $h$-periodic functions. Prove

\medpagebreak {\bf Lemma 3.3.} \ {\it If $\im \xi>-\pi+h/2$, then
\begin{equation}  \sigma\,(2i\xi-\pi)\,m\,(z,\,\xi)=
\sigma\,(2i\xi-3\pi)\,[\alpha\,m\,(z,\,\xi+i\pi)+
\beta\,m\,(z,\,\xi+2i\pi)],\label{3.28}\end{equation} where
$\alpha$ and $\beta$ are entire functions of $\xi$  and entire and
$h$-periodic in $z$. }

\demo As equation \eqref{3.0} and \eqref{3.1} are equivalent, then
the space of solutions of \eqref{3.1} is also a two dimensional
module over the ring of $h$-periodic functions $\IK$, and,
moreover, for any two solutions $f$ and $g$ of \eqref{3.1}, the
expression
$$\{f(z),g(z)\}=f(z+h)g(z)-f(z)g(z+h)$$
is $h$-periodic, and $f$ and $g$ are linearly independent over
$\IK$ iff $\{f(z),g(z)\}\not\equiv 0$. We call $\{f(z),g(z)\}$ the
wronskian of $f$ and $g$.
\smallpagebreak Assuming additionally to the hypothesis of the
lemma that $\xi$ satisfies the condition
\begin{equation} 1/\sigma\,(2i\xi-3\pi)\ne
0,\label{3.29}\end{equation} e.i. that $$\xi\ne -i\pi+ih/2+i\pi
l+ihk, \quad  l,k=0,\,1,\,2,\,\dots\,,$$
we  check that $e^{i\pi z/h}\,m\,(z,\,\xi+i\pi)$ and
$m\,(z,\,\xi+2i\pi) $  form a base in the space of entire
solutions of \eqref{3.1}. For this, we calculate their wronskian.
We shall write $f\sim g$ if $f$ differs from $g$ by an entire in
$\xi$ factor having no zeros.

\smallpagebreak First, we study the asymptotics of the wronskian
for $z\to\pm i\infty$. They can be easily calculated by means of
the asymptotics of $m$. Assuming that $b_0(\xi+i\pi)$ and
$a_0(\xi+2i\pi)$ are nonzero, we get \begin{equation} \{e^{i\pi
z/h}\,m\,   (z,\,\xi+i\pi),\,m\,(z,\,\xi+2i\pi)\}\sim
b_0(\xi+i\pi)\,a_0(\xi+2i\pi)\,(1+o\,(1)).\label{3.30}\end{equation}
$$z\to+i\infty$$ But, in fact, formula \eqref{3.21} shows that
$a_0(\xi)\ne 0$ for all $\xi\in\C$, and formula \eqref{3.22}
implies that $b_0(\xi+i\pi)\ne 0$ under the condition
\eqref{3.29}. This justifies \eqref{3.30}. Similarly, one shows
that
\begin{equation} \{e^{i\pi
z/h}\,m(z,\,\xi+i\pi),\,m\,(z,\,\xi+2i\pi)\}\sim
d_0(\xi+i\pi)\,c_0(\xi+2i\pi)(1+o\,(1)).\label{3.31}\end{equation}
$$z\to-i\infty$$ Since the wronskian is an entire $h$-periodic
function of $z$, \eqref{3.30} and  \eqref{3.31} imply that it is
in fact independent of $z$, $$\{e^{i\pi
z/h}\,m\,(z,\,\xi+i\pi),\,m\,(z,\,\xi+2i\pi)\} \sim
d_0(\xi+i\pi)\,c_0(\xi+2i\pi).$$

\smallpagebreak Furthermore, taking into account \eqref{3.23} and
\eqref{3.24}, we see that
\begin{equation} \{e^{i\pi z/h}\,m\,(z,\,\xi+i\pi),\,m\,(z,\,\xi+2i\pi)\}\sim
1/\sigma\,(2i\xi-3\pi),\label{3.32}\end{equation} and thus, under
the condition \eqref{3.29}, \ $e^{i\pi z/h}\,m\,(z,\,\xi+i\pi)$
and $m\,(z,\,\xi+2i\pi) $ are basis solutions. Therefore, in this
case, $m\,(z,\,\xi)$ is their linear combination,
$$m\,(z,\,\xi)=\tilde\alpha\,e^{i\pi z/h}\,m\,(z,\,\xi+i\pi)+
\tilde\beta\,m\,(z,\,\xi+2i\pi)$$ with the coefficients
$$\tilde\alpha= \{m\,(z,\,\xi),\,m\,(z,\,\xi+2i\pi)\}/ \{e^{i\pi
z/h}\,m\,(z,\,\xi+i\pi),\,m\,(z,\,\xi+2i\pi)\},$$ $$\tilde\beta=
\{e^{i\pi z/h}\,m\,(z,\,\xi+i\pi),\,m\,(z,\,\xi)\}/ \{e^{i\pi
z/h}\,m\,(z,\,\xi+i\pi),\,m\,(z,\,\xi+2i\pi)\}.$$ One can
calculate the wronskians in these formulae in the same manner as
above, which gives $$ \{m\,(z,\,\xi),\,m\,(z,\,\xi+2i\pi)\}
=w_1(\xi)+w_2(\xi)\,e^{\dsize-\frac{2\pi i}h\,z},$$ $$\{e^{i\pi
z/h}\,m\,(z,\,\xi+i\pi),\,m\,(z,\,\xi)\}=w_3(\xi),$$ $$w_j\sim
1/\sigma\,(2i\xi-\pi),\quad j=1,\,2,\,3.$$ This and \eqref{3.32}
imply the desired relation \eqref{3.28} under the condition
\eqref{3.29}. But, then it is valid for any $\xi$  in the
half-plane $\im \xi>-\pi+h/2$ as an equality of two meromorphic
functions.

\qed

\medpagebreak {\bf 2.} \ Now, we note that the integral
\eqref{3.2} remains analytic in $\xi$ and entire in $z$ while the
pole $p_0-\pi-h=i\xi-\pi-h/2$ of the function $v$ remains in the
left half-plane of the complex plane, and its pole
$-p_0+\pi+h=-i\xi+\pi+h/2$ remains in the right half-plane. Thus,
$m$ is analytic in $\xi$ if $\im \xi>-\pi-h/2$.

\smallpagebreak
The above result implies that  Lemma 3.3 remains valid
for all $\xi$ in the half-plane $\im \xi>-\pi-h/2$.

\smallpagebreak
{\bf 3.} \
Consider the function
$$\Phi\,(z,\,\xi)=\sigma\,(2i\xi-\pi)\,m\,(z,\,\xi).$$
Since the poles of $\sigma\,(2i\xi-\pi)$ are in the half-plane $\im \xi\ge
h/2$, \
$\Phi$ is analytic in $\xi$ in the strip $-\pi-h/2<\im \xi<h/2$.
Lemma 3.3 immediately implies

\medpagebreak {\bf Lemma 3.4.} \ {\it The function $\Phi$ can be
analytically continued in the half-plane $\im\xi<h/2$.} \demo The
right hand side in \eqref{3.28} is analytic  in $\xi$ if
$-2\pi-h/2<\im \xi<-\pi+h/2$. Therefore, $\Phi$ is analytic in the
strip $-2\pi-h/2<\im \xi<h/2$. By means of \eqref{3.11}, one can
rewrite formula \eqref{3.28} in the form $$\Phi\,(z,\,\xi)=
\alpha\,(z,\,\xi)\,\Phi\,(z,\,\xi+i\pi)+
\beta\,(z,\,\xi)\,\left(1+e^{\dsize
\frac{2\pi}{h}\,(\xi+2i\pi)}\right) \Phi\,(z,\,\xi+2i\pi),$$ and
this relation allows to continue $\Phi$ analytically from the
strip $-2\pi-h/2<\im \xi<h/2$ in the whole half-plane
$\im\xi<h/2$. \qed

\medpagebreak
The lemma implies that $m\,(z,\,\xi)$ can be meromorphically  continued
on the whole complex plane. We shall not distinguish between $m$ and its
meromorphic continuation.
The $m$ is analytic in $\xi$ in the upper half-plane of the complex
plane, and  the product $\sigma\,(2i\xi-\pi)\,m\,(z,\,\xi)$ is analytic
in the half-plane $\im\xi\le 0$.

\smallpagebreak {\bf 4.} \ Remind that  $m$ is entire in $z$ for
$\im \xi>-\pi+h/2$. As the meromorphic continuation of $m$ into
the whole plane of $\xi$ was obtained by \eqref{3.28}, $m$ is
entire in $z$ for all $\xi$ different from the zeros of
$\sigma(2i\xi-\pi)$, i.e. outside the poles of $m$.

\smallpagebreak {\bf 5.} \ Since $m\,(z+h,\,\xi)$, $m\,(z,\,\xi)$
and $m\,(z-h,\,\xi)$ are meromorphic in $\xi$, equation
\eqref{3.1} remains satisfied for all $\xi\in\C$ as a relation for
 meromorphic in $\xi$ functions.

\medpagebreak {\bf 6.} \ To complete the proof of the theorem, we
have to check only its statement ({\it ii}). Consider the
canonical Bloch solutions of \eqref{3.1} described by \eqref{3.26}
-- \eqref{3.27}. One can represent
$\Phi\,(z,\,\xi)=\sigma\,(2i\xi-\pi)\,m\,(z,\,\xi)$ by their
linear combination with $h$-periodic coefficients. In a vicinity
$\C_+$ of $+i\infty$,
$$\Phi\,(z,\,\xi)=A\,(z,\,\xi)\,f_1(z,\,\xi)+B\,(z,\,\xi)\,f_2(z,\,\xi),\quad
z\in \C_+.$$ The coefficients $ A\,(.,\xi),\,\,B\,(.,\xi)$ can be
expressed in terms of the wronskians of $\Phi$ and $f_{1,2}$:
$$A=\{\Phi,\,f_2\}/\{f_1,f_2\},\quad
B=\{f_1,\,\Phi\}/\{f_1,f_2\}.$$ Thus, they are analytic in $\xi$
in the same part of the complex plane as $\Phi$. Also, all the
Fourier coefficients of $A$ and $B$  are analytic in $\xi$ in the
same domain as $\Phi$, and, therefore are described by the same
analytic formulae as for $\im \xi>-\pi+h/2$, so everywhere outside
the poles of $\sigma$. The asymptotics \eqref{3.19} --
\eqref{3.20} show that $$A\to a_0(\xi)\,\sigma\,(2i\xi-\pi),\quad
B\to b_0(\xi)\,\sigma\,(2i\xi-\pi),$$ $$\im \xi>-\pi+h/2,\quad
z\to+i\infty,$$ outside the poles of $\sigma\,(2i\xi-\pi)$. So,
the zero-th Fourier coefficients of $A$ and $B$ are equal
respectively to $a_0\sigma$ and $b_0\sigma$ for all $\xi$ outside
the poles of $\sigma(2i\xi-\pi)$, and all the Fourier coefficients
with the negative indices are zero on the whole complex plane of
$\xi$. This implies the statement of the Lemma concerning the
representation \eqref{3.19}. The applicability of \eqref{3.20} can
be extended in the same way.\qed
\subsection{ Estimates of the solution $m$}
Here, we just list some estimates for $m\,(z)$ and $m'(z)$
immediately following from the asymptotic representations
\eqref{3.19} and \eqref{3.20}. These estimates will be used in the
next section. One has

\begin{equation} |m\,(z)|\le C\, P\,(z), \quad  |m'(z)|\le C\,
(1+|y|)\,P\,(z),\label{3.33}\end{equation} where
$$P\,(z)=e^{\dsize -|y|/2}\,\left\{\begin{array}{cl} e^{\dsize
|x-\pi-\varphi|\,y/h}, & y\ge0,\\ e^{\dsize
|x+\varphi|\,|y|/h}\,e^{\dsize -\pi|y|/h},& y\le 0,
\end{array}\right.$$
where $x=\re z$, $y=\im z$ and  $\varphi=\im\xi$.
These estimates are uniform in $\re z$ if $|\re z|$ is bounded by a fixed
constant.
\section{ Minimal solutions of a perturbed auxiliary equation}
In this section, we consider the equation
\begin{equation} \psi\,(z+h)+\psi\,(z-h)+2e^{\xi}\cos
z\,\psi\,(z)=w\,(z)\psi\,(z), \label{4.1}\end{equation} assuming
that $\xi$ is complex number, and $w$ is a meromorphic function
satisfying the estimate \begin{equation} |w\,(z)|\le C\,e^{\dsize
(1-\mu)\,|y|},\quad 0<\mu, \quad |y|>Y,\label{4.2}\end{equation}
for sufficiently big $Y$. Here, as below, in this section, we use
the notations $$x=\re z,\quad y=\im z.$$ Our aim is to prove the
existence of a meromorphic solution of \eqref{4.2} having the
asymptotics representations of the same form as ones of the
minimal solution $m$ of equation \eqref{3.1}.

\smallpagebreak Remind that $m$ has poles if
$\sigma(2i\xi-\pi)=0$, e.i. if $$2i\xi=2\pi+h+2hl+2\pi m,\quad
l,\,m=0,\,1,\,2,\,\dots\,.$$ Let $$\Sigma=\{\pm (2\pi+h +2hl+2\pi
m),\,\,\,l,m=0,\,1,\,2,\,\dots\,\}.$$ In this section, we assume
that \begin{equation} 2i\xi\not\in\Sigma.\label{4.3}\end{equation}

\smallpagebreak One can try to invert the operator being in the
left-hand side \eqref{4.1}. As we shall see, this leads to an
integral equation of the form
\begin{equation} \psi\,(z)=m\,(z)+\int_\gamma
\kappa\,(z,\,\zeta)\psi\,(\zeta)\,d\zeta,\quad z\in \gamma.
\label{4.4}\end{equation} Here $ \gamma$ is a vertical curve, and
$\kappa$ is constructed in terms of $m$ and $\tilde m$ two
linearly independent solutions of \eqref{3.1}, \begin{equation}
\kappa\,(z,\,\zeta)= \frac 1{2ih}\,\theta\,(z,\,\zeta)\, \frac{
        [\,m(z)\,\tilde m\,(\zeta)-m\,(\zeta)\,\tilde m\,(z)\,]}
        {\{m,\,\tilde m\}}\,w\,(\zeta),
        \label{4.5}\end{equation}
where
\begin{equation} \theta\,(z,\,\zeta)=\ctg\frac{\pi(\zeta-z)}h\,+i,\label{4.6}\end{equation}
and $\{m,\,\tilde m\}=m(z+h) \tilde m(z)-m(z)\tilde m(z+h)$ is the
{\it wronskian} of $m$ and $\tilde m$. This integral equation is
our main tool. Note that the kernel $\kappa$ can be considered as
a difference analog of the resolvent kernel arising in the theory
of differential equations: now, instead of the canonical
$\theta$-function, $\theta(y-\eta)$, which equals to zero if
$y>\eta$ and to $1$ if $y<\eta$, one encounters the function
\eqref{4.5}.

First, we shall study the integral equation \eqref{4.4}, and then
we shall study its solutions and, in particular, check that they
satisfy \eqref{4.1}. The analysis of the integral equation is
quite similar to one we have carried out in the case of Harper
equation $\psi\,(z+h)+\psi\,(z-h)+2\cos z\,\psi\,(z)=E\psi\,(z)$,
\ $E=\Const$, in \cite{BF6}. So, we describe in details only its
new elements.

The main results concerning \eqref{4.1} are formulated in
subsection 4.4.
\subsection{ Operator $K$} Now, we are going to describe precisely
the construction of the integral operator from \eqref{4.4} (the
solutions $m$ and $\tilde m$, the integration contour), to
describe the functional space and  to   discuss properties of this
operator.
\subsubsection{The solutions $m$ and $\tilde m$}
Let $m$ be the minimal solution of \eqref{3.1} constructed in
Section 3, put \begin{equation} \tilde
m\,(z,\,\xi)=\overline{m\,(\overline{z},\overline{\xi})}.
\label{4.7}\end{equation} Here, we indicated explicitly the
dependence of $m$ on $\xi$. Together with $m$, \ $\tilde m$ is an
entire in $z$ solution of \eqref{3.1}.

Remind that $m$ is analytic in $\xi$ outside the set of zeros of
$\sigma\,(2i\xi-\pi)$. Thus, both $m$ and $\tilde m$ are analytic
under the condition \eqref{4.3}.

The new solution has the asymptotics $$\hskip-4cm \tilde m\,(z)=
(\cc{c_0(\cc{\xi})}+o\,(1))\, e^{\dsize -\frac
i{2h}\,(z-\pi+i\xi)^2+\frac{iz}2}+$$ \begin{equation} \hskip2cm
+e^{\dsize \frac{2\pi i}h\,z} (\cc{d_0\,(\cc{\xi})}+o\,(1))\,
e^{\dsize \frac i{2h}\,(z-\pi+i\xi)^2+\frac {iz}2},
\label{4.8}\end{equation} $$z\to+i\infty,$$ and $$\hskip-4cm
\tilde m\,(z)= (\cc{a_0(\cc{\xi})}+o\,(1))\, e^{\dsize -\frac
i{2h}\,(z-\pi-i\xi)^2-\frac{iz}2}+$$ \begin{equation} \hskip2cm
(\cc{b_0(\cc{\xi})}+o\,(1))\, e^{\dsize \frac
i{2h}\,(z-\pi-i\xi)^2-\frac {iz}2}, \label{4.9}\end{equation} $$
z\to-i\infty.$$ Here, $a_0$, $b_0$, $c_0$ and $d_0$ are the
coefficients from the asymptotic formulae for $m$.
\subsubsection{The wronskian of $m$ and $\tilde
m$} \label{wronskian} When proving Lemma 3.3, we have already
discussed the space of solutions of equation \eqref{3.1}, the
linear independence and the wronskians of its solutions. We
calculate the wronskian of $m$ and $\tilde m$ as we have
calculated wronskians in the proof of Lemma 3.3. We recall that
the wronskian is $h$-periodic and get its asymptotics for $z\to\pm
i\infty$ (using the asymptotics of $m$ and $\tilde m$). This leads
to the result
$$\left\{m\,(z),\,\tilde
m\,(z)\right\}=e^{\xi}\,\cc{c_0(\cc{\xi})}\,a_0(\xi). $$
This and  formulae \eqref{3.21} -- \eqref{3.23} imply
\begin{equation} \left\{m\,(z),\,\tilde m\,(z)\right\}= -4\pi
ih\,e^{\dsize \xi/2}.\label{4.10}\end{equation}
Note that this result means that the solutions $m$ and $\tilde m$
are linearly independent over the ring of $h$-periodic in $z$
functions.
\subsubsection{The integration contour} Let
$\varphi=\im \xi$. The contour $\gamma$ is a strictly vertical
curve which does not pass through any pole of $w$, comes from
$-i\infty$ along the line
\begin{equation} x=-\varphi,\quad ,\label{4.11}\end{equation} and
goes to $+i\infty$ along the line
\begin{equation} x=\pi+\varphi.\label{4.12}\end{equation}

\smallpagebreak To motivate this specific choice of the asymptotes
of $\gamma$, note that  the estimates \eqref{3.33} show that,
$|m|$ is minimal along $\gamma$:
\begin{equation} |m\,(z)|\le C\,e^{\dsize -|y|/2}\,\frac1{p_0(z)},
\quad z\in\gamma,\label{4.13}\end{equation}

\begin{equation} p_0(z)=\left\{\begin{array}{cl} 1, & y>0,\\ e^{\dsize \pi|y|/h},
& y\le 0.\end{array}\right.\label{4.14}\end{equation}

\smallpagebreak Note also that  \begin{equation} |\tilde
m\,(z)|\le C\,e^{\dsize -|y|/2}\, p_0(z),\quad \quad z\in\gamma.
\label{4.15}\end{equation}
\subsubsection{ The functional space} We define
the operator $K$ as the integral operator with the kernel
\eqref{4.5} acting in the space $L_2\,(\gamma,\,p)$, where $p$ is
the weight reflecting as the behavior of $m$ and $\tilde m$ along
$\gamma$, so the one of $w$, see \eqref{4.2}, \begin{equation}
p\,(z)=e^{\dsize (1-\mu)|y|}\,p_0^2(z).\label{4.16}\end{equation}
\subsubsection{ Compactness of the operator $K$}
The kernel of the operator $K$ is quickly decaying along the
contour $\gamma$: using the estimates \eqref{3.32} for $m$ and
$m'$, one can show that \begin{equation}
p^{1/2}(z)\,|\kappa\,(z,\zeta)|\,p^{-1/2}(\zeta)\le
C\,(1+|\eta|)\,
e^{\dsize-\frac{\mu}2\,|y|-\frac{\mu}2\,|\eta|},\quad \eta=\im
\zeta, \quad z,\,\zeta\in\gamma. \label{4.17}\end{equation} The
estimate \eqref{4.17} implies that {\it the operator $K$ is a
Hilbert-Schmidt operator in $L_2(\gamma,p)$}.
\subsection{ Solutions of equation \eqref{4.1} in a
vicinity of the curve $\gamma$} Note that, $m\in L_2(\gamma,p)$.
Therefore, at least, one of  the equations
\begin{equation} \psi=m+K\psi\quad{\rm and}\quad
\psi=K\psi,\label{4.18}\end{equation}
has a nontrivial solution $\psi\in L_2(\gamma,p)$.   In the
sequel, $\psi$ denotes this nontrivial solution. If both the
equations have nontrivial solutions from $L_2(\gamma,p)$, then
$\psi$ is one of them. Show that $\psi$ can be analytically
continued into a vicinity of $\gamma$ up to a solution of equation
\eqref{4.1}.
\subsubsection{ Analytic continuation}  One has

\bigpagebreak
{\bf Proposition 4.1.} \ {\it There exists a positive $\delta$, such
that $\psi$ can be analytically
continued in the horizontal $(h+\delta)$-vicinity of $\gamma$.}

\medpagebreak \demo For $\zeta\in \gamma$, the kernel
$\kappa\,(z,\,\zeta)$ is analytic in $z$ in the horizontal
$h$-vicinity $S_h$ of $\gamma$. The integral, representing
$K\psi$, converges uniformly in $z$ being in any compact subset of
$S_h$. Therefore, the function $K\psi$ can be analytically
continued in $S_h$. But then, $\psi$, being a solution of
\eqref{4.18}, can be also analytically continued in $S_h$.

Let $\delta$ be the horizontal distance from $\gamma$ to the
closest to it pole of the function $w$. If $\delta<h$, then having
proved that $\psi$ is analytic in the horizontal $h$-vicinity of
$\gamma$, one can deform the integration contour in the formula
for $K\psi$ inside its horizontal $\delta$-vicinity, and check
that, in fact, $\psi$ can be analytically continued in the
horizontal $h+\delta$-vicinity. If $\delta>h$, then one deforms
the contour inside its horizontal $h$-vicinity  to check that
$\psi$ is analytic in $2h$-vicinity of $\gamma$ and so on.

\qed

\medpagebreak For brevity, we shall not distinguish $\psi$ and its
analytic continuation.
\subsubsection{$\psi$ and equation
\eqref{4.1}}\label{residues} Note that the horizontal width of the
horizontal $(h+\delta)$-vicinity $S_{h+\delta}$ of $\gamma$ is
bigger than $2h$. Show that $\psi$ satisfies equation \eqref{4.1}
for $z-h,\,z,\,z+h\,\in S_{h+\delta}$.

Let $({\mathcal H}_0\,f)\,(z)=f\,(z+h)+f\,(z-h)+2e^{\xi}\,\cos
z\,f\,(z)$. Then, ${\mathcal H}_0\,\psi={\mathcal H}_0\,K\psi$,
but $$({\mathcal H}_0\,K\, \psi)\,(z)= 2\pi
i\Res_{\zeta=z}\,\kappa\,(z+h,\,\zeta)\,\psi\,(\zeta)=$$
$$=\frac{(m\,(z+h)\,\tilde m\,(z)-m\,(z)\,\tilde m\,(z+h))}
{\{m,\,\tilde m\}}\,w\,(z)\,\psi\,(z)=w\,(z)\,\psi\,(z).$$ Thus,
we come to

\bigpagebreak {\bf Proposition 4.2.} \ {\it The solution $\psi$
satisfies equation \eqref{4.1} in the strip $S_{h+\delta}$}.
\subsubsection{ Asymptotics of $\psi$ for $z\to
+i\infty$} Let us describe the asymptotic behavior of $\psi$ in a
vicinity of $\gamma$. We begin by proving

\bigpagebreak {\bf Proposition 4.3.} \ {\it Fix positive
$\varepsilon_1$ and $\varepsilon_2$ so that $\varepsilon_1<2h$ and
$\varepsilon_2<\min\{\mu,\,2\pi/h\}$. In the horizontal
$\varepsilon_1$-vicinity of $\gamma$, the solution $\psi$ admits
the asymptotic representation
\begin{equation} \psi\,(z)=m\,(z)\,(A+A_1(z))+\tilde m\,(z)\,(B+B_1(z)),\quad
z\to+i\infty, \label{4.19}\end{equation} where \begin{equation}
A=\delta(\psi)+\frac1{Wh}\, \int_{\gamma}\tilde
m\,(\zeta)\,w\,(\zeta)\,\psi\,(\zeta)\,d\zeta,\quad B= -\frac
1{Wh}\,
\int_{\gamma}m\,(\zeta)\,w\,(\zeta)\,\psi\,(\zeta)\,d\zeta,
\label{4.20}\end{equation} and  the functions $A_1$ and $B_1$
satisfy the estimates
\begin{equation} |A_1(z)|,\,|B_1(z)|\le C\,e^{-\varepsilon_2\,y},\quad
y>Y,\label{4.21}\end{equation} for sufficiently big  $Y$, finally,
$W=\{m,\,\tilde m\}$, and
\begin{equation}\label{4.20a}
 \delta(\psi)=\left\{\begin{array}{cc} 1,&\quad {\rm if }\quad
\psi=m+K\psi,\\ 0, & \quad {\rm if  }\quad
\psi=K\psi.\end{array}\right.
\end{equation}
} \smallpagebreak {\bf Remark. } \ { \it Note that the form of the
above asymptotic representation depends on the integral equation
for $\psi$.}

\bigpagebreak The proof of this statement is almost the same as
the proof of Proposition 3.6 from \cite{BF6}. Here, we just
outline the main idea of the proof. The function $\psi$ satisfies
one of the equations from \eqref{4.18}. Thus, on the contour
$\gamma$, one can estimate it by estimating the right hand side of
the equation. Using the inclusion $\psi\in L_2(\gamma,p)$ and the
estimate \eqref{4.17}, one obtains
$$|\psi\,(z)|\le\Const e^{-|y/2|}\,/p_0(z).$$
The integrals
$$ \int_\gamma m\,(\zeta)\,w\,(\zeta)\,\psi\,(\zeta)\,d\zeta,\quad
\int_\gamma \tilde m\,(\zeta)\,w\,(\zeta)\,\psi\,(\zeta)\,d\zeta
$$
quickly converge as  $\zeta\to+i\infty$, and one can obtain the
asymptotics of $K\psi\,\,(z)$ for $z\to+i\infty$ inside the
horizontal $h$-vicinity of $\gamma$ by replacing in the kernel of
$K$ the function $\theta\,(z,\,\zeta)$ \ with $2i$. This leads to
the answer $$\psi\,(z)\sim A\,m\,(z)+B\,\tilde m\,(z).$$ We omit
the elementary estimates leading to \eqref{4.21}. To prove
\eqref{4.19} outside the horizontal $h$-vicinity of $\gamma$,  one
has to write down a convenient representation for the analytic
continuation of $K\psi$ outside $\gamma$. For example, if $z$ is
to the right of $\gamma$, and the horizontal distance between $z$
and $\gamma$ is between $h$ and $2h$, one can use the formula
$$(K\psi)\,(z)=2\pi i\,\Res_{\zeta=z-h}
\kappa\,(z,\,\zeta)\,\psi\,(\zeta)+
\int_\gamma\kappa\,(z,\,\zeta)\,\psi\,(\zeta)\,d\zeta.$$

\qed

\bigpagebreak Using the representation \eqref{4.19} for $\psi$ and
the asymptotics \eqref{3.19} --\eqref{3.20} for $m$, one finally
obtains the asymptotics of $\psi$ for $z$ being in the horizontal
$\varepsilon_1$-vicinity of $\gamma$:
$$\hskip-4cm \psi\,(z)= (a+o\,(1))\,e^{\dsize \frac
i{2h}\,(z-\pi+i\xi)^2+\frac{iz}2}+$$
\begin{equation} \hskip2cm +(b+o\,(1))\,e^{\dsize -\frac
i{2h}\,(z-\pi+i\xi)^2+\frac i2\,z}, \quad
z\to+i\infty,\label{4.22}\end{equation} with \begin{equation}
a=A\,a_0(\xi),\quad b=A\,b_0\,(\xi)+
B\overline{c_0(\overline{\xi})},\label{4.23}\end{equation}
where $a_0$, $b_0$ and $c_0$ are the coefficients from the
asymptotic representations \eqref{3.19} -- \eqref{3.20} for
$m\,(z)$.
\subsubsection{ Asymptotics of $\psi$ for $z\to
-i\infty$}  Describing the behavior of $\psi$ for $z\to-i\infty$,
one proves first  the representation
\begin{equation}
\psi\,(z)=(C+C_1(z))\,m\,(z)+ (D+D_1(z))\,e^{\dsize -\frac{2\pi
i}h\,z}\,\tilde m\,(z),\quad z\in
S_{\varepsilon_1}(\gamma),\quad\im z\to-\infty.
\label{4.24}\end{equation}
Here, $C_1$ and $D_1$ are functions satisfying the estimates of
the form \eqref{4.21}, and $C$ and $D$ are constant coefficients
given by $$C=\delta(\psi),\quad D=-\frac 2{Wh}\, \int_\gamma
e^{+2\pi i \zeta}m\,(\zeta)\,w\,(\zeta)\,\psi\,(\zeta)\, d\zeta.$$
\smallpagebreak {\bf Remark.} \ { \it Let us emphasize that, now,
the asymptotic representation strongly depends on the integral
equation for $\psi$: if $\psi$ is a solution of the inhomogeneous
equation, then $C=1$; if $\psi$ is a solution of the homogeneous
equation, then $C=0$.}

\medpagebreak The representation \eqref{4.24} implies that, in the
horizontal $\varepsilon_1$-vicinity of $\gamma$, $$\hskip-4cm
\psi\,(z)= (c+o\,(1))\,e^{\dsize \frac
i{2h}\,(z-\pi-i\xi)^2-\frac{iz}2}+$$ \begin{equation} \hskip2cm
\,e^{\dsize -\frac{2\pi i}h\,z}\, (d+o\,(1))\,e^{\dsize -\frac
i{2h}\,(z-\pi-i\xi)^2-\frac i2\,z}, \quad
z\to-i\infty,\label{4.25}\end{equation} with $$c=C\,c_0(\xi),\quad
d=C\,d_0(\xi)+D\,\overline{a_0(\overline{\xi})}.$$ where $a_0$,
$c_0$ and $d_0$ are the coefficients from the asymptotic
representations \eqref{3.19} -- \eqref{3.20} for $m\,(z)$.
\subsection{ Solution $\psi$ outside the vicinity of $\gamma$}
\subsubsection{Analyticity near $\pm i\infty$} Being
analytic in the horizontal $h+\delta$-vicinity of $\gamma$, the
function $\psi$ can be meromorphically continued into the whole
complex plane just by means of equation \eqref{4.1}. Since $w$ is
analytic in some vicinities $\C_\pm$ of $\pm i\infty$, \ $\psi$ is
also analytic there.
\subsubsection{Asymptotics in $\C_\pm$}
Discuss the
asymptotics of $\psi$ as $y\to\pm\infty$ for arbitrary fixed $x$.
Begin with the case of $z\to+i\infty$. As for equation
\eqref{3.1}, see subsection 3.4.3, one can construct for
\eqref{4.1} the canonical basis solutions $f_{1,2}$ analytic in a
vicinity $\C_+$ of $+i\infty$.  This can be carried in the same
way as in the proof of Theorem 1.1. One obtains
\begin{equation} f_{1,2}(z)= e^{\dsize
\pm\frac{i}{2h}\,(z-\pi+i\xi)^2+\frac i2\,z}\, (1+o\,(1)),\quad
y\to+i\infty.\label{4.26}\end{equation} In $\C_+$, one can
represent $\psi$ by their linear combination with $h$-periodic
analytic coefficients
\begin{equation} \psi\,(z)=A\,(z)\,f_1(z)+B\,(z)\,f_2(z).\label{4.27}\end{equation}
Now, we recall that the asymptotic representation \eqref{4.22} is
valid in the $\varepsilon_1$-vicinity of $\gamma$ and that as
$\varepsilon_1$ one can be choose any number from $(0,2h)$. In
particular, this representation is valid in a horizontal vicinity
of the horizontal length bigger then $h$, i.e. the period of $A$
and $B$. Therefore, comparing this representation with
\eqref{4.22}, we see that the periodic coefficients $A$ and $B$
are bounded as $z\to+i\infty$, and
\begin{equation} A\,(z)\to a,\quad B\,(z)\to b,\quad z\to
+i\infty.\label{4.28}\end{equation}
Now, as $A$ and $B$ are $h$-periodic, \eqref{4.26} -- \eqref{4.28}
imply that the asymptotic representation \eqref{4.22} remains
valid and uniform  in $\re z$ if $\im z\to +i\infty$ and $|\re z|$
is bounded by any fixed constant.

Reasoning in the same way one proves that the asymptotic
representation \eqref{4.25} for $\psi$ as $y\to-i\infty$ remains
valid and (locally) uniform  in $\re z$ too.
\subsection{ Results of the section}
The main results of this section can be formulated in the following form.

\bigpagebreak {\bf Proposition 4.4.} \ {\it Let
$2i\xi\not\in\Sigma$. Let also $w$ be a meromorphic function
satisfying the estimate \eqref{4.2}. Assume that $\gamma$ is a
strictly vertical curve not passing through any pole of $w$.  Then
equation \eqref{4.1} has a meromorphic solution $\psi\,(z)$ with
the following properties:

{\bf 1.} \ $\psi$ is analytic in $z$ in some vicinities $\C_\pm$
of $\pm i \infty$;

{\bf 2.} \ there is  a positive $\delta$ such that $\psi$ has no
poles in the horizontal $(h+\delta)$-vicinity of $\gamma$;

{\bf 3.} \ $\psi$  admits the asymptotic representations of the
form $$\hskip-4cm \psi\,(z,\,\xi)= (a+o\,(1))\, e^{\dsize \frac
i{2h}\,(z-\pi+i\xi)^2+\frac{iz}2}+$$ \begin{equation} \hskip2cm
+(b+o\,(1))\, e^{\dsize -\frac i{2h}\,(z-\pi+i\xi)^2+\frac i2\,z},
\quad y\to+\infty,\label{4.29}\end{equation} and $$\hskip-4cm
\psi\,(z,\,\xi)= (c+o\,(1))\, e^{\dsize \frac
i{2h}\,(z-\pi-i\xi)^2-\frac{iz}2}+$$ \begin{equation} \hskip2cm
\,e^{\dsize -\frac{2\pi i}h\,z}\, (d+o\,(1))\, e^{\dsize -\frac
i{2h}\,(z-\pi-i\xi)^2-\frac i2\,z}, \quad
y\to-\infty,\label{4.30}\end{equation} where $a$, $b$, $c$, and
$d$ are independent of $z$. The error estimates are uniform in $x$
if $|x|$ is bounded by a constant.}
\smallpagebreak {\bf Remark.} \ {\it The solution $\psi$ satisfies
one of the integral equations \eqref{4.18}, if it satisfies the
homogenous equation, then the coefficient $c$ in the asymptotic
representation \eqref{4.30} is zero.}
\section{ Minimal entire solutions of equation \eqref{1.1}}
In this section, we construct the minimal entire solutions of the
matrix equation \eqref{1.1} and prove Theorem 1.2.
\subsection{ The plan}
We start with the equation \eqref{1.5} for the first component of
a vector solution of \eqref{1.1}. We represent $\psi_1$ in the
form \begin{equation}
\psi_1(z)=t\,(z)\,f_0(z),\label{5.1}\end{equation} where $t$
satisfies the equation
\begin{equation} t\,(z+h)=\rho\,(z)\,t\,(z-h).\label{5.2}\end{equation} This
transforms \eqref{1.5} to
\begin{equation} f_0(z+h)+f_0(z-h)=v_1\,(z)\,f_0(z)\label{5.3}\end{equation} with
\begin{equation} v_1\,(z)=\frac{t\,(z)}{t\,(z+h)}\,v\,(z).\label{5.4}\end{equation}
We find  a solution $t$ of \eqref{5.2} such that
$t\,(z)\,/\,t\,(z+h)$ tends to some nonzero constants as $z\to\pm
i\infty$.

\smallpagebreak Then, we accept assumption \eqref{1.20}. In this
case, $$v_1\,(z)\sim {v_1}_-\,e^{in z}+{v_1}_+\,e^{-in z},\quad
z\to\pm i\infty, $$ and, by a linear change of the variable, one
transforms \eqref{5.3} to equation \eqref{4.1} investigated in the
previous section. This allows to construct some analytic solution
$\psi_1$ of equation \eqref{1.5}. This function can be considered
as the first component of a vector solution $\psi$ of the original
matrix equation \eqref{1.1}. The second component can be
reconstructed by formula \eqref{1.6}.

\smallpagebreak To be sure that  $\psi$ is really entire in $z$,
we check that  it is analytic in a sufficiently wide horizontal
vicinity a vertical curve $\gamma$. Really, since $\det
M\,(z)\equiv1$, the formulae \begin{equation}
\psi\,(z+h)=M\,(z)\,\psi\,(z),\quad
\psi\,(z-h)=M^{-1}(z-h)\,\psi\,(z)\label{5.5}\end{equation} show
that, in this case, $\psi$ can be continued up to an entire
function of $z$.
\subsection{ Meromorphic solution of equation
\eqref{5.2}} Let  $\gamma$ be a strictly vertical curve. Here, we
construct a meromorphic solution of \eqref{5.2} analytic in a
vicinity of $\gamma$. We shall use the following notations:
\begin{itemize}
\item we denote by
$\gamma+2\pi$ the curve obtained of $\gamma$ by the
$2\pi$-translation;
\item we denote by $S_\gamma$ the strip
bounded by $\gamma$ and $\gamma+2\pi$ so that $\gamma\subset
S_\gamma$ and $\{\gamma+2\pi\}\cap S_\gamma=\emptyset$;
\item we let $n_\pm=n_\pm(b)$ and $N=n_++n_-$;
\item we denote the zeros of $b(z)$ situated in $S_\gamma$ by
$z_l$, \ $l=1,2,\dots N$.
\end{itemize}
\medpagebreak {\bf Proposition 5.1.} \ {\it Let $\gamma$ be a
strictly vertical curve  not passing through any point where
either $b\,(z)=0$ or $b\,(z-h)=0$. There exists a meromorphic
solution $t$ of equation \eqref{5.2} and a positive number
$\delta$ such that

\point{i} $t$ is analytic in the horizontal $(h+\delta)$-vicinity
of $\gamma$; \point{ii} if $z_0$ is a zero of $b\,(z)$ situated in
this vicinity to the left of $\gamma$, then $t\,(z_0)=0$; in
addition, the multiplicity of the zero of $t$ equals to the
multiplicity of the zero of $b\,(z)$ at $z=z_0$;
\point{iii} the function $v_1$ defined by \eqref{5.4} is analytic
in the horizontal $\delta$-vicinity of $\gamma$;
\point{iv} the function $t$ admits the asymptotic representations
\begin{equation} t\,(z)=e^{\dsize in_-z/2}\, (1+\tau_-(z)),\quad
\im z\to-\infty,\label{5.6}\end{equation} \begin{equation}
t\,(z)=t_\infty\,e^{\dsize -in_+z/2}\, (1+\tau_+(z)),\quad \im
z\to+\infty, \label{5.7}\end{equation} where
\begin{equation} t_\infty=\exp\,\left(\frac i2\sum_{l=1}^{n_+ + n_-}z_l\,-i\pi
N/2+ihN/4 \right),\label{5.8}\end{equation} and $\tau_\pm$ are
functions satisfying the estimates \begin{equation}
|\tau_\pm(z)|\le C\,e^{-\mu|\im z|}, \label{5.9}\end{equation}
where $\mu$ is a positive number. This formulae are uniform in
$\re z$ if $|\re z|$ is bounded by a fixed constant.}

\medpagebreak \demo {\bf 1.} \ Construct a solution of \eqref{5.2}
in terms of the $\sigma$-function introduced in subsection 3.2.
The function $\rho\,(z)=b\,(z)/b\,(z-h)$ can be represented in the
form \begin{equation}  \rho\,(z)=e^{\dsize in_-
h}\,\prod_{l=1}^{N} \rho_l(z),\quad \rho_l(z)=\frac{1-e^{
-i(z-z_l)}} {1-e^{ -i(z-z_l-h)}}.\label{5.10}\end{equation}
Substituting this representation in equation \eqref{5.2}, and
comparing the result with \eqref{3.4}, one immediately  finds out
that the function \begin{equation} t\,(z)=e^{\dsize in_-z/2}\,
\prod_{l=1}^{N}\,t_l(z),\quad
t_l(z)=\frac{\sigma\,(z+\pi-z_l)}{\sigma\,(z+\pi-z_l-h)},\label{5.11}\end{equation}
is a solution of  \eqref{5.2}. Clearly, it is meromorphic in $z$.

\medpagebreak {\bf 2.} \ To prove the proposition, we have to
recall some properties of the $\sigma$-function. First, we note
that the zeros of the $\sigma$-function are situated  at the
points $\pi+h+2hl+2\pi m$, \ $l,m\in\N\cup\{0\}$, and that its
poles are at the points $-\pi-h-2hl-2\pi m$, \
$l,m\in\N\cup\{0\}$. Secondly, we remind that as the pole at
$z=-\pi-h$ so the zero at $z=\pi+h$ are simple. Moreover, if $2\pi
m <2h$, \ $m\in \N$, then all the poles and zeros situated at the
points $z=-\pi-h-2\pi l$, and $z=\pi+h+2\pi l$, \
$l=0,\,\dots\,m$, are simple.

\smallpagebreak The ratio
$\frac{\sigma\,(z+\pi)}{\sigma\,(z+\pi-h)}$ has poles only at the
points $$z=2h+2hl+2\pi m,\quad {\rm and}\quad  -2\pi-h-2h l-2\pi
m,\quad l,m\in \N\cup\{0\}.$$

\smallpagebreak The zeros of this ratio are at the points
$$z=h+2hl+2\pi m,\quad {\rm and}\quad  -2\pi-2h l-2\pi m,\quad
l,m\in \N\cup\{0\},$$ the zeros at $z=h$ and at $z=-2\pi$ are
simple. Moreover, if $2\pi m <2h$, \ $m\in \N$, then all the zeros
situated at the points $z=h+2\pi l$ and $z=-2\pi-2\pi l$, \
$l=0,\,\dots\,m$, are simple.

\smallpagebreak
Now, we shall show that the described properties of the ratio
$\frac{\sigma\,(z+\pi)}{\sigma\,(z+\pi-h)}$
imply the first three statements of the proposition.

\medpagebreak
{\bf 3.} \
Show that there is a positive number $\delta$ such that
the constructed $t$ is analytic in the horizontal $h+\delta$-vicinity of
$\gamma$.

\smallpagebreak
Denote by  $\delta_1$ the horizontal distance between the set of the points
$z_l$, \ $l=1,\,\dots\,,N$, and the curve $2\pi+\gamma$.
The poles of the factor $t_l(z)$ closest to $\gamma$ are at the points
$z_l+2h$ and $z_l-2\pi-h$. So, it is analytic in  the
horizontal $h+\delta$-vicinity of $\gamma$, where
$$\delta=\min\{\delta_1,\,h\}.$$ This implies the first statement of the
proposition.

\medpagebreak {\bf 4.} Let us prove the second statement. Let
$z_0$ be a zero of $b\,(z)$ situated to the left of the curve
$\gamma$ inside the above horizontal $h+\delta$-vicinity of
$\gamma$. Show that  $t\,(z_0)=0$.

\smallpagebreak All the factors $t_l$ are analytic in the vicinity
of $\gamma$. Inside this vicinity, to the left of $\gamma$, the
factor $t_l$ equals to zero only at the points $z_l-2\pi m$,
$m=1,\,2,\,\dots$. This proves the first part of the second
statement. Prove the second part.

\smallpagebreak The choice of $\delta$ implies that a point of the
form $z_l-2\pi m$, \ $m=2,3,4,\dots$, can be situated in the
horizontal $(h+\delta)$-vicinity of $\gamma$ only if $2\pi m<h$.
So, all the zeros of $t_l$ being in this vicinity to the left of
$\gamma$ are simple. In result, the multiplicity of the zero of
$t\,(z)$ at any of these points being in the vicinity equals to
the number of the factors $t_l$ which are equal to $0$ there, e.i.
to the multiplicity of the zero of $b$ at this point.

\medpagebreak {\bf 5.} The function $v_1\,(z)$ defined by
\eqref{5.4} is clearly meromorphic. It can be also represented in
the form
\begin{equation} v_1(z)=\frac{\dsize t\,(z)}{\dsize t\,(z+h)}\,a(z)+ \frac{\dsize
t\,(z)}{\dsize t\,(z-h)}\,d(z-h).\label{5.12}\end{equation}
Formula \eqref{5.12} shows that the poles of $v_1$ situated in the
above horizontal vicinity of $\gamma$ can be only  at the points
where $t\,(z+h)=0$ or (and) $t\,(z-h)=0$, e.i. at the points
$$z=z_l+2hl+2\pi m,\quad {\rm and}\quad  z=z_l-2\pi+h-2h l-2\pi
m,\quad l,m\in \N\cup\{0\}.$$ Among them, only the points
$$z_l+2\pi m\quad {\rm and}\quad z_l-2\pi m+h,\quad
m=0,\,1,\,\dots,$$ can be in the above vicinity. So, they can be
only the points where either $b\,(z)=0$ or $b\,(z-h)=0$. However,
the curve $\gamma$ does not pass through any of these points.
Thus, there is a positive $\delta_2$ such that the function $v_1$
is analytic in the horizontal $\delta_2$-vicinity of $\gamma$. If
this $\delta_2$ is smaller than the previously defined $\delta$,
we simply redefine $\delta=\delta_2$.

\medpagebreak {\bf 6.} \ The asymptotics of $t$ for $z\to\pm i
\infty$ follow immediately from formula \eqref{5.11} and the
asymptotics \eqref{3.9} and \eqref{3.10} of the $\sigma$-function.

\qed
\subsection{ Entire solutions of \eqref{1.1}}
Here, following the plan described above, we shall construct a set
of entire solutions of \eqref{1.1}.

\bigpagebreak {\bf 1.}  \ Let $\gamma$ be a strictly vertical
curve do not passing through any point where either $b(z)=0$ or
$b(z-h)=0$, and let $t$ be the  solution of equation \eqref{5.2}
described in Proposition 5.1. Let us study in more details the
function $v_1$ defined by \eqref{5.4}.

\smallpagebreak By Proposition 5.1, $v_1$ is analytic in the
$\delta$-vicinity of $\gamma$. In the case of \eqref{1.20}, the
asymptotic representations for $t$ imply that \begin{equation}
v_1(z)=-e^{\dsize i(nz+\phi_-+\pi)}-
e^{\dsize-i(nz+\phi_++\pi)}+w\,(z),\label{5.13}\end{equation}
where \begin{equation} \phi_-=-n_-(b)\,h/2-i\ln v_- ,\quad
\phi_+=-n_+(b)\,h/2+i\ln v_+ ,\label{5.14}\end{equation} and $w$
is a function satisfying the estimate $$|w\,(z)|\le C\,e^{\dsize
(n-\mu)\,|\im z|},\quad |\im z|>Y,$$ where $\mu$ is the same
number as in \eqref{5.9}, and $Y$ is a sufficiently big positive
number. This estimate is uniform in $\re z$ if $|\re z|$ is
bounded by a constant.

\medpagebreak {\bf 2.} \ Let \begin{equation} \phi=(\phi_+ +
\phi_-)/2,\quad 2i\xi=\phi_+-\phi_-.\label{5.15}\end{equation}
Equation \eqref{5.3} can be rewritten in the form:
\begin{equation} f_0(z+h)+f_0(z-h)+2e^\xi\,\cos
(nz+\phi+\pi)\,f_0(z)=w\,(z)\,f_0(z). \label{5.16}\end{equation}

\smallpagebreak The change of the variable \begin{equation} z_1=
nz+\phi+\pi,\label{5.17}\end{equation} transforms \eqref{5.16} to
the equation \begin{equation}
\psi\,(z_1+h_1)+\psi\,(z_1-h_1)+2e^\xi\,\cos (z_1)\,\psi\,(z_1)=
w_1(z_1)\,\psi\,(z_1), \label{5.18}\end{equation} where
$$\psi\,(z_1)=f_0(z),\quad w_1(z_1)=w\,(z),\quad h_1=nh.$$ This
equation was investigated in the previous section.

\medpagebreak {\bf 3.} \ Now, we can construct a solution of
\eqref{5.16} by means of Proposition 4.4.

\smallpagebreak Check that the coefficients of \eqref{5.18}
satisfy its assumptions. The estimate of $w\,(z)$ implies that
$w_1$ satisfies an estimate of the form \eqref{4.2} as a function
of $z_1$. Note that the branches of the logarithms in \eqref{5.14}
are fixed modulo $2\pi$. To ensure the assumption on $\xi$, we fix
them so that the parameters $\phi_\pm$ satisfy condition
\eqref{1.18}.

\smallpagebreak To apply the proposition, it rests to choose a
vertical curve mentioned in it. We denote this curve $\gamma_1$
and choose it so that it be the image under the transformation
$z\mapsto z_1$ of the curve $\gamma$ used for constructing of the
function $t$. To apply the proposition, we have to check that
$\gamma_1$ does not not pass by any pole of $w$. But this follows
from the statement (iii) of Proposition 5.1.

\smallpagebreak Now, we construct the solution $\psi$ of
\eqref{5.18} by means of Proposition 4.4. Note that this solution
is analytic in the horizontal $(h_1+\delta_1)$-vicinity of
$\gamma_1$,  where $\delta_1$ is a positive number. This number is
determined by the horizontal distance between the poles of $w_1$
and the curve $\gamma_1$.  We can assume that $\delta_1=n\delta$ (
if $\delta_1>n\delta$, we redefine it letting $\delta_1=n\delta$;
if $\delta_1<n\delta$, we redefine $\delta$).

\smallpagebreak Indicating explicitly the dependence  of $\psi$ on
$\xi$ and $h$, we let
\begin{equation} f_0(z)=\psi\,(n_0z+\phi+\pi,\,\xi,\,nh).\label{5.19}\end{equation}
The function $f_0$ is a solution of \eqref{5.16}. It is analytic
in the horizontal $h+\delta$-vicinity of $\gamma$.

\medpagebreak {\bf 4.} \ The function $\psi_1=t\,f_0$ is a
solution of equation \eqref{1.5}, and, thus, can be considered as
the first component of a vector solution $\Psi$ of the  matrix
equation \eqref{1.1}. Its second component can be reconstructed by
formula \eqref{1.6}.

\medpagebreak
{\bf 5.} \ Let us discuss the analytic properties of $\psi_1$.

\smallpagebreak First, note that the asymptotics of $f_0$
(following from  Proposition 4.3), and the asymptotics for $t$
described by \eqref{5.6} and \eqref{5.7} imply the representations
$$\hskip-4cm \psi_1(z)= (A_0+o\,(1))\, e^{\dsize \frac
i{2hn}\,(nz+\phi_+)^2+\frac{iz}2\,(n-n_+)}+ $$ \begin{equation}
+(B_0+o\,(1))\, e^{\dsize -\frac
i{2hn}\,(nz+\phi_+)^2+\frac{iz}2\,(n-n_+)},\quad \im z\to+\infty,
\label{5.20}\end{equation} $$\hskip-4cm \psi_1(z)= (C_0+o\,(1))\,
e^{\dsize \frac i{2hn}\,(nz+\phi_-)^2-\frac{iz}2\,(n-n_-)}+$$
\begin{equation} +\,e^{\dsize -\frac{2\pi i}h\,z}\, (D_1+o\,(1))\,
e^{\dsize -\frac i{2hn}\,(nz+\phi_-)^2-\frac{iz}2\,(n-n_-)}, \quad
\im z\to-\infty,\label{5.21}\end{equation} where
$n_\pm=n_\pm(b)$, and $A_0$, $B_0$, $C_0$, and $D_1$ are
coefficients independent of $z$.

\smallpagebreak
Secondly, note that since both $t$ and $f_0$ are analytic in the horizontal
$h+\delta$-vicinity of $\gamma$, the function $\psi_1$ is also analytic there.

\medpagebreak {\bf 6.} \ Discuss the analyticity of the second
component of the vector $\Psi=\left(\begin{array}{c} \psi_1\\
\psi_2\end{array}\right)$. For any $\epsilon>0$, we denote by
$L_\gamma(\epsilon)$ the part of the horizontal
$\epsilon$-vicinity of $\gamma$ situated to the left of $\gamma$.
Clearly, $\psi_1(z+h)-a\,(z)\,\psi_1(z)$ is analytic in
$L_\gamma(h+\delta)$. So, if \ $b$ has no zeros in this domain,
$\psi_2(z)=(\psi_1(z+h)-a\,(z)\,\psi_1(z))/b\,(z)$ is also
analytic there.

\medpagebreak {\bf 7.} \  As the horizontal length of
$L_\gamma(h+\delta)$ is greater then $h$ and $\Psi$ is analytic
here, then we can continue $\Psi$ in the whole complex plane by
means of equation \eqref{1.1}: we continue $\Psi$ to the right
just by means this equation itself, and we continue it to the left
by the formula $\Psi(z-h)=M^{-1}(z-h) \Psi(z)$ (remind that $\det
M\equiv 1$).

\medpagebreak We have proved

\medpagebreak {\bf Proposition 5.2.} \ {\it Assume that
$b\not\equiv 0$ and that $v\,(z)$ satisfies condition
\eqref{1.20}. By means of formulae \eqref{5.14}, choose some
numbers $\phi_\pm$ satisfying \eqref{1.18}. Let   $\gamma$ be a
strictly vertical curve not passing through any point where either
$b\,(z)=0$ or $b\,(z-h)=0$. If the curve can be drown so that
there would be no zeros of the function $b$ in the domain
$L_\gamma(h)$, then equation \eqref{1.1} has an entire solution
$\Psi$ with the first component admitting the asymptotic
representations \eqref{5.20} -- \eqref{5.21}.}

\smallpagebreak In fact, the solution $\Psi$ described in
Proposition 5.2 is one of the main personages of this   paper. In
subsection 5.5, we shall see that it is a {\it minimal} entire
solution of equation \eqref{1.1}.
\subsection{ Modified approach}
In the previous subsection, we have constructed a (minimal) entire
solution of equation\eqref{1.1} under a hypothesis on the geometry
of  the set of zeros of $b(z)$. We were
 assuming that there is a strictly vertical curve
$\gamma$ such that $b(z)$ and $b(z-h)$ have no zeros  on $\gamma$
and $b(z)$ has no zeros in $L_\gamma(h)$, i.e. in the left part
the horizontal $h$-vicinity of $\gamma$. Now, we shall remove this
geometric condition. We begin with the case where $b$ has only one
simple zero $z_0\in L_\gamma(h)$.  Let us outline the idea.
\subsubsection{The idea} Any solution of \eqref{1.5} can be considered
as the first component $\psi_1$ of a vector solution $\Psi=
\left(\begin{array}{c} \psi_1\\\psi_2 \end{array}\right)$ of
equation \eqref{1.1}.  Its second component can be recovered by
\eqref{1.6}.  If $\psi_1$ is analytic in the horizontal
$(h+\delta)$-vicinity of a vertical curve $\gamma$ and such that
\begin{equation} \psi_1(z_0+h)-a\,(z_0)\,\psi_1(z_0)=0,\label{5.22}\end{equation}
then the corresponding vector solution $\Psi$ is analytic at least
in $L_\gamma(h+\delta)$ and, so, it can be analytically continued
up to an entire function just by means of \eqref{1.1}.

In the case of Proposition 5.2, we have constructed (minimal)
entire solutions of \eqref{1.1} in terms of a solution $f_0$ of
equation \eqref{5.3} analytic in the horizontal
$(h+\delta)$-vicinity of $\gamma$, using the formula
$\psi_1(z)=t\,(z)\,f_0(z)$. Now, by Proposition 5.1 (statement
(ii)),
\begin{equation}
t\,(z_0)=0.\label{5.23}\end{equation}
So, $\psi_1=t\,f_0$ appears to be  analytic in the horizontal
$(h+\delta)$-vicinity of $\gamma$  even if $f_0$ is replaced by a
solution of \eqref{5.3}  having there one (simple) pole situated
at $z_0$. For the "old" analytic at $z_0$ solution $f_0$, the
condition \eqref{5.22} is not satisfied. The idea is to satisfy
this condition by considering instead of the "old" $f_0$ linear
combinations of the "old"  $f_0$ and a "new" solution of
\eqref{5.3}  having one simple pole at $z_0$.

Of course, the meromorphic solutions of \eqref{5.3} we are looking
for have to possed asymptotic representations for $z\to\pm
i\infty$ of the same analytic structure as $f_0$ so that $\psi_1$
would have the asymptotics of the form \eqref{5.20} --
\eqref{5.21}.

To construct such solutions of \eqref{5.3}, we shall consider a
solution $m\,(z,\,z_0)$ of the model equation \eqref{3.1}
possessing the analogous properties. Then, we shall plug it into
the integral equation~\eqref{4.4} in the place of $m\,(z)$ (the
integral operator remains the same as before) and study its
solutions.
\subsubsection{The solution $m(z,z_0)$}
We shall use the variable $z$ of the input matrix equation
\eqref{1.1}. Let $\phi$ and $\xi$ be as in \eqref{5.15}. Consider
the equation (compare it with \eqref{5.16})
\begin{equation} g(z+h)+g(z-h)+2e^{\xi}\,\cos (z_1)\,g(z)=0,\quad
z_1=nz+\phi+\pi.\label{5.24}\end{equation} The functions
$m(z_1,\,\xi,\,h_1)$ and $\tilde m(z_1,\,\xi,\,h_1)$, \ $h_1=nh$,
are solutions of \eqref{5.24}. Here, we have indicated explicitly
the dependence of these functions on $h$ and $\xi$.

\smallpagebreak
Let
\begin{equation}
m\,(z,\,z_0)=\theta\,(z,\,z_0)\,\frac{[\,m(z_1)\,\tilde m\,({z_0}_1+h_1)-
m\,({z_0}_1+h_1)\,\tilde m\,(z_1)\,]}
        {\{m,\,\tilde m\}},\label{5.25}\end{equation}
where ${\{m,\,\tilde m\}}$ is the wronskian of $m(z)$ and $\tilde
m(z)$, see subsection \ref{wronskian},
$${z_0}_1=nz_0+\phi+\pi,$$
and, as before, $$\theta\,(z,\,z_0)=\ctg\frac{\pi(z_0-z)}h\,+i.$$

\smallpagebreak Since the wronskian ${\{m,\,\tilde m\}}$ is
constant, and $\theta\,(z,z_0)$ is $h$-periodic in $z$, the
function $m\,(z,\,z_0)$ satisfies equation \eqref{5.24}.

\smallpagebreak Let us discuss analytic properties of $m(z,z_0)$.
It is analytic outside the points $z=z_0+lh$, \ $l\in \Z$. As
$[\,m({z_0}_1)\,\tilde m\,({z_0}_1+h_1)- m\,({z_0}_1+h_1)\,\tilde
m\,({z_0}_1)\,]={\{m,\,\tilde m\}}$, the solution $m(z,z_0)$ has a
simple pole at $z=z_0$,
$$\Res_{z=z_0} m\,(z,z_0)=\frac h\pi.$$
At the point $z=z_0+h$, the numerator in \eqref{5.25} equals to
zero, and, thus, in fact, $m(z,z_0)$ is analytic here. So, for
some positive $\delta$, \ $m\,(z,\,z_0)$ is analytic in the
horizontal $h+\delta$-vicinity of $\gamma$ if $z\ne z_0$.

\smallpagebreak The asymptotics of $m(z)$ and $\tilde m(z)$ for
$z\to\pm i\infty$  imply the asymptotic representations for
$m(z,z_0)$:

 $$\hskip-4cm m\,(z,z_0)=
 A\,(z_0)\,e^{\dsize  \frac
i{2hn}\,(nz+\phi_+)^2+\frac{inz}2}\,(1+o\,(1))+$$
\begin{equation} \hskip1cm +B\,(z_0)\,e^{\dsize -\frac
i{2hn}\,(nz+\phi_+)^2+\frac{inz}2}\,(1+o\,(1)), \quad \im
z\to+i\infty,\label{5.26}\end{equation}

$$\hskip-2cm m\,(z,z_0)=e^{\dsize -2\pi iz/h}\,\left(
C\,(z_0)\,e^{\dsize \frac
i{2hn}\,(nz+\phi_-)^2-\frac{inz}2}\,(1+o\,(1))+\right.$$
\begin{equation} \hskip1cm +\left. D\,(z_0)\,e^{\dsize -\frac
i{2hn}\,(nz+\phi_-)^2-\frac{inz}2}\,(1+o\,(1))\right), \quad \im
z\to-i\infty,\label{5.27}\end{equation} where $A$, $B$, $C$ and
$D$ are independent of $z$.
\subsubsection{The integral operator} To construct the
meromorphic $f_0$, we shall again use the integral operator from
\eqref{4.4}, but now, it will be convenient to write it in terms
of the variable of the input equation \eqref{1.1}. So, now,
$\gamma$ is a strictly vertical curve going to $+ i\infty$ along
the line $\re z=-\re \phi_+\,/\,n$, and from $- i\infty$ along the
line $\re z=-(\pi+\re \phi_-)\,/\,n$. We define the weight
$P(z)=e^{(n-\mu)|y|}\,p_0^2(z_1)$, where $y=\im z$, \  $\mu$ is
the constant from the estimate for $w\,(z)$ from \eqref{5.13}, and
$p_0$ is defined by \eqref{4.14}. Both the functions $m(.)$ and
$m\,(.,\,z_0)$ belong to $L_2\,(\gamma,\,P)$.

\medpagebreak The kernel of the integral operator takes the form
\begin{equation} \kappa\,(z,\,\zeta)=
\frac 1{2ih}\,\theta\,(z,\,\zeta)\, \frac{
        [\,m\,(z_1)\,\tilde m\,(\zeta_1)-
           m\,(\zeta_1)\,\tilde m\,(z_1)\,]}
        {\{m,\,\tilde m\}}\,w\,(\zeta),\quad \zeta_1=n\zeta+\varphi+\pi.
\label{5.28}\end{equation} Here, $w$ is the same as in
\eqref{5.18}. The kernel satisfies the estimate
\begin{equation}
P^{1/2}(z)\,|\kappa\,(z,\zeta)|\,P^{-1/2}(\zeta)\le
C\,(1+|\eta|)\,
e^{\dsize-\frac{\mu}2\,|y|-\frac{\mu}2\,|\eta|},\quad y=\im
z,\quad \eta=\im \zeta, \quad z,\,\zeta\in\gamma.
\label{5.4.17}\end{equation}
The integral operator $K$ with the
kernel \eqref{5.28} is compact in the space $L_2\,(\gamma,\,P)$.
\subsubsection{Auxiliary construction}  To construct the
meromorfic $f_0$, we use an  operator equation slightly different
from \eqref{4.18}. Before writing down this equation, we shall
prove an auxiliary statement motivating this equation. Note that,
since $\det M\equiv1$, the equality $b\,(z_0)=0$ implies that
$a\,(z_0)\ne0$. We put
\begin{equation} s\,(z_0)=\frac\pi{h}\,\frac{t\,(z_0+h)}{t'\,(z_0)\,a\,(z_0)}.
\label{5.29}\end{equation}

\smallpagebreak
One has

\medpagebreak {\bf Lemma 5.3.} \ {\it Assume that there exists a
function $f\in L_2(\gamma,\,P)$ which can be analytically
continued in a vicinity of the point $z_0+h$, and satisfies the
relation
\begin{equation} f(z)=Kf(z)+s\,(z_0)\,m\,(z,\,z_0)\,f(z_0+h)+\hat m\,(z),\quad
z\in \gamma,\label{5.30}\end{equation} where either $\hat
m\,(z)\equiv m\,(z_1)$ or $\hat m\,(z)\equiv 0$. Let
$\psi_1(z)=t\,(z)\,f(z)$. This function has the following
properties:

\point{i} it is a solution of \eqref{1.5} analytic in the
horizontal $(h+\delta)$-vicinity of $\gamma$, where $\delta$ is a
positive number;

\point{ii} $\psi_1$ satisfies the relation \eqref{5.22};

\point{iii}  for $\im z\to\pm i\infty$, it admits the asymptotic
representations of the form \eqref{5.20} -- \eqref{5.21}.}

\medpagebreak {\bf Remark.} \ {\it As we shall see from the proof,
if the function $f$ satisfying \eqref{5.30} exists, then it
satisfies equation \eqref{5.3}. It is the needed meromorphic
solution of this equation. Furthermore, the function $\psi_1= tf$
is the first component of the desired (minimal) entire vector
solution of equation \eqref{1.1}.}

\medpagebreak \demo {\bf a.} \ Since $f\in L_2(\gamma,\,P)$, and
the kerenel of $K$ satisfies the estimate \eqref{5.4.17}, then the
term $Kf$ is analytic in the horizontal $h$-vicinity of $\gamma$.
Therefore, the right hand side of \eqref{5.30} is analytic in this
vicinity except the point $z=z_0$. This implies that $f$ itself is
also analytic there. At the point $z=z_0$, this function has a
simple pole,
\begin{equation} {\rm
res}_{z=z_0}\,f(z)=\frac{h}{\pi}\,s\,(z_0)\,f(z_0+h).\label{5.31}\end{equation}

\smallpagebreak Let $\delta_1$ be the horizontal distance from
$z_0$ to $\gamma$. Deforming the integration contour in the
definition of $K$, one checks that, in fact, $f$ is analytic in
the horizontal $(h+\delta_1)$-vicinity of $\gamma$ without the
point $z=z_0$.

\smallpagebreak {\bf b.} Using a calculation analogous to one from
subsection \ref{residues}, by means of the residue theorem, we
check that the relation \eqref{5.30} implies that, in the above
vicinity of $\gamma$, \ $f$ satisfies the equation \eqref{5.16}
which is just a different form of writing of \eqref{5.3}.

\smallpagebreak {\bf c.} \ Remind that $t$ is analytic in a
horizontal $(h+\delta)$-vicinity of $\gamma$, and $t\,(z_0)=0$. If
$\delta>\delta_1$, we redefine it  letting $\delta=\delta_1$. Let
$\psi_1=t\,f$. This function is analytic in the whole horizontal
$(h+\delta)$-vicinity of $\gamma$. Since $f$ satisfies
\eqref{5.3}, $\psi_1$ is a solution of equation \eqref{1.5}.

\smallpagebreak {\bf d.} \ Check that $\psi_1$ satisfies the
relation \eqref{5.22}. Using \eqref{5.31} and \eqref{5.29}, we see
that $$\psi_1(z_0)=t'\,(z_0)\,{\rm res}_{z=z_0}\,f(z)=
\frac{t\,(z_0+h)}{a\,(z_0)}\,\,f(z_0+h)=\frac{1}{a\,(z_0)}\,\psi_1(z_0+h).$$

\smallpagebreak {\bf e.} \ Finally, applying the methods of
section 4 to equation \eqref{5.30}, one obtains the asymptotics of
$f$ for $z\to\pm i\infty$ which leads to (iii).

\qed
\subsubsection{Operator equation}  Now, we turn to constructing
a function $f$ satisfying the hypothesis of Lemma 5.3. Since
\eqref{5.30} can not be used directly as an equation for a
function from $L_2(\gamma,\,P)$, we change the functional space to
${\mathcal H}=L_2\,(\gamma,\,P)\oplus \C$. For $F\in{\mathcal H}$,
we use the notation $$F=\left(\begin{array}{c} f\\
s\end{array}\right),\quad f\in L_2\,(\gamma,\,P),\quad s\in\C.$$
Then we define the operator ${\mathcal K}$ acting in ${\mathcal
H}$ by the formula $${\mathcal K}\,F=\left(\begin{array}{cc} Kf &
m\,(z,z_0)\,s\\ Kf(z_0+h) & m\,(z_0+h,z_0)\,s\end{array}\right).$$
For  $f\in L_2\,(\gamma,\,P)$, the function $Kf(z) $ is analytic
in the horizontal $h$-vicinity of $\gamma$, and thus the number
$K\,f(z_0+h)$ is well defined.

\smallpagebreak One can easily check that the application
$f\longrightarrow K\,f(z_0+h)$ is a linear bounded operator from
$L_2\,(\gamma,\,P)$ to $\C$, which implies that ${\mathcal K}$ is
a compact operator together with $K$.

\smallpagebreak The new integral equation is \begin{equation}
{\mathcal K}\,F+ F_0=F,\quad F_0(z)=\left(\begin{array}{c} \hat
m\,(z)
\\ \hat m\,(z_0+h)\end{array}\right). \label{5.32}\end{equation}
We let $\hat m\,(z)\equiv m\,(z_1)$ if $1\not\in{\rm
spec}\,{\mathcal K}$ and $\hat m\,(z)\equiv 0$ otherwise.

\smallpagebreak Clearly, \eqref{5.32} always has a nontrivial
solution. For $F$ being a solution of this equation,
\begin{equation} f\,(z)=Kf\,(z)+m\,(z,z_0)\,s+\hat
m\,(z),\label{5.33}\end{equation}
\begin{equation} s=Kf\,(z_0+h)+m\,(z_0+h,z_0)\,s+\hat
m\,(z_0+h).\label{5.34}\end{equation} Discuss properties of the
function $f$.
\subsubsection{The solution of the operator
equation  and the hypothesis of Lemma 5.3} Since $f\in
L_2(\gamma,\,P)$, it can be analytically continued in a vicinity
of $z_0+h$ just by means of \eqref{5.33}.

\smallpagebreak Letting in \eqref{5.33} $z=z_0+h$ and comparing
the result with \eqref{5.34}, we see that $$s=f\,(z_0+h).$$
Substituting this expression for $s$ in \eqref{5.33}, we come to
\eqref{5.30}.

\smallpagebreak Thus, we see that $f$ satisfies the assumptions of
Lemma 5.3. This means that we have constructed the desired
(minimal) entire solution.
\subsubsection{Several zeros of $b$}  We have proved the existence
of the minimal solution in the case where $b$ has only one simple
zero in the domain $L_\gamma(h)$. The case of several zeros (even
multiple ones) can be treated similarly. In particular, if,  in
$L_\gamma(h)$, there are $J$  simple zeros of $b$, then instead of
\eqref{5.30} one considers the relation $$f(z)=Kf(z)+\sum
_{j=1}^{J}s\,(z_j)\,m\,(z,\,z_j)\,f(z_j+h)+ \hat m\,(z),\quad z\in
\gamma,$$ and instead of ${\mathcal H}$ one has to introduce the
space $L_2\,(\gamma,\,P)\oplus \C^J$. We omit the details and
formulate the result:

\medpagebreak {\bf Proposition 5.4.} \ {\it  Assume that
$b\not\equiv 0$ and that $v\,(z)$ satisfies condition
\eqref{1.20}. By means of formulae \eqref{5.14}, choose some
numbers $\phi_\pm$ satisfying \eqref{1.18}. Equation \eqref{1.1}
has an entire solution $\Psi$  with the first component $\psi_1$
admitting the asymptotic representations \eqref{5.20} --
\eqref{5.21}.}
\medpagebreak {\bf Remark.} \ {\it Note that  if $1\in{\rm
spec}\,{\mathcal K}$ then, in the asymptotic representation
\eqref{5.21} for $\psi_1$, the coefficient $C_0$ is zero}
\subsection{ Minimal entire solutions} \ {\bf 1.} Now, let us
discuss in more detail the behavior of the
solution described in Proposition 5.4  for $z\to \pm i\infty$. We
shall use the canonical basis solutions $f_{1,2}$ and $g_{1,2}$
analytic in some vicinities $\C_+$ and $\C_-$ of $+i\infty$ and
$-i\infty$ correspondingly, see Theorems 1a and 1b.

\smallpagebreak Remind that any entire solution $\psi$ of
\eqref{1.1} admits the representations \eqref{1.19}. Consider
these representations for the solution $\Psi$ described in
Proposition 5.4.

\smallpagebreak Remind that the asymptotics \eqref{1.10} of the
basis Bloch solutions $f_{1,2}$ contain parameter $\phi_+$. This
parameter is defined by \eqref{1.11} modulo $2\pi$. The choice of
the value of the parameter is a choice of two particular solutions
$f_{1,2}$.

\smallpagebreak Formulae \eqref{1.11} and \eqref{5.14} coincide,
and thus, $\phi_+$ from \eqref{1.10} can be chosen equal to
$\phi_+$ in \eqref{5.20}. In this case, comparing \eqref{5.20} and
\eqref{1.10}, we see that in a vicinity of $+i\infty$,  \ $\Psi$
admits the first of the representations \eqref{1.19} with $A$ and
$B$ bounded as $z\to+i\infty$.

\smallpagebreak Choosing the canonical basis $g_{1,2}$ in a
vicinity of $-i\infty$ with the same values of $\phi_-$ as in
\eqref{5.21}, one also sees that $\Psi$ admits the second of the
representations \eqref{1.19} with $C$ and $D$ bounded as
$z\to-i\infty$, and that, moreover, $D\,(z)\to0$ as
$z\to-i\infty$.

\smallpagebreak Now, we recall the definition of the minimal
entire solutions of \eqref{1.1}, and see that the solution $\Psi$
is one of them. It is the solution $\psi_D$.

\bigpagebreak {\bf 2.} \ The proofs of the existence of the other
three  minimal solutions are similar to the above one. So, we
shall discuss only a place  where they essentially differ. This is
the integral equation being the starting point for the analysis of
Section 4. Consider for example the case of the solution for which
$B\,(z)\to 0$ as $z\to+i\infty$. For this solution, equation
\eqref{4.4} has to be replaced by \begin{equation}
\psi\,(z)=m\,(2\pi-z)+ \int_{\{2\pi-\gamma\}}
\kappa\,(2\pi-z,\,2\pi-\zeta)\psi\,(\zeta)\,d\zeta.
\label{5.35}\end{equation} with the same kernel $\kappa$ as in
Section 4. The contour $\{2\pi-\gamma\}$ is obtained of the
contour $\gamma$ from Section 4 by the mapping $z\to 2\pi-z$. As
in Section 4, constructing a minimal solution of \eqref{1.1}, we
have used  as the first term in the right hand side of the
integral equation the minimal solution of equation \eqref{3.1}
having the same form of the asymptotics for $z\to\pm i\infty$ as
the solution we construct. The analysis of equation \eqref{5.35}
is similar to one of equation \eqref{4.4}.

\medpagebreak {\bf 3.} We have described how one can construct
four minimal entire solutions corresponding to the canonical bases
with the given parameters $\phi_\pm$. The only condition on these
to numbers is given by \eqref{1.18}. In fact, it means that the
canonical bases have to be consistent. This remark finishes the
proof of our Theorem 1.2.
\section{Asymptotic coefficients of minimal entire solutions
and the basic properties of these solutions and of the
corresponding monodromy matrices} Choose two canonical bases
$f_{1,2}$ and $g_{1,2}$. Consider the corresponding minimal
solutions. Assume that these solutions exist and that their
asymptotic coefficients are non zero. In this section, we study
some basic properties of the minimal solutions and of the
corresponding monodromy matrices. We shall use the notations from
the section "Asymptotic coefficients" of the introduction.
\subsection{ Wronskians of the minimal solutions}
Here, we study the wronskians of the minimal entire
solutions and  check Proposition 1.3.
One has

\medpagebreak
{\bf Lemma 6.1.} \ {\it The wronskian of any two of the four
minimal solutions corresponding to the
bases $f_{1,2}$ and $g_{1,2}$ is independent of $z$.}

\demo
Consider the minimal solutions $\psi_D$ and
$\psi_B$.
Study their wronskian. Remind that it is an $h$-periodic entire function.

\smallpagebreak In a vicinity of $+i\infty$,
$$\{\psi_D(z),\,\psi_B(z)\}=
\{A^{(D)}(z)f_1(z)+B^{(D)}(z)\,f_2(z),
\,A^{(B)}(z)\,f_1(z)+B^{(B)}(z)\,f_2(z)\}=$$
$$=w_f\left(A^{(D)}\,(z)\,B^{(B)}(z)-A^{(B)}\,(z)\,B^{(D)}(z)\right).$$
Here, we have marked by the letters $B$ and $D$ the functions
$A\,(z)$ and $B\,(z)$ from the representations \eqref{1.19} for
the solutions $\psi_D$ and $\psi_B$. The obtained formula for
their wronskian implies that \begin{equation}
\{\psi_D(z),\,\psi_B(z)\}\to -w_f\,A_B\,B_D,\quad
z\to+i\infty.\label{6.1}\end{equation}

\smallpagebreak Similarly, one proves that
\begin{equation} \{\psi_D(z),\,\psi_B(z)\}\to w_g\, C_D\,D_B,\quad
z\to-i\infty.\label{6.2}\end{equation}

\smallpagebreak
Thus, the wronskian is  a bounded entire function, and, therefore, it is
independent of $z$.

The cases of the other pairs of  the minimal solutions can be treated in
the same way.
\qed

\medpagebreak Asymptotics \eqref{6.1} and \eqref{6.2} imply the
formulae
\begin{equation} \{\psi_D,\,\psi_B\}=w_g\,C_D\,
D_B=-w_f\,A_B\,B_D.\label{6.3}\end{equation} In fact, this is two
of the formulae described in Proposition 1.3. The other formulae
can be derived similarly.
\subsection{ Uniqueness of the minimal entire solutions}
Here, we deduce from Proposition 1.3 its Corollary 1.5. It suffices to check

\medpagebreak {\bf Lemma  6.2.} \ {\it Consider two of the minimal
solutions corresponding to the  canonical bases $f_{1,2}$ and
$g_{1,2}$. If their wronskian is nonzero, then each of these
solutions is unique up to an independent of $z$ factor.}

\demo
Consider the case of the solutions
$\psi_D$  and $\psi_B$. The other cases can be treated similarly.

\smallpagebreak Let $\psi$ be a minimal solution for which, as for
$\psi_D$, $D\,(z)$ tends to zero as $z\to-i\infty$. Show that
$\psi\,(z)=\Const \psi_D(z)$. By Lemma  6.1, the wronskian
$w=\{\psi_D,\psi_B\}$ is independent of $z$, and by the
assumption, it is nonzero. Thus, $\psi_D$  and $\psi_B$  form a
base in the space of entire solutions of \eqref{1.1}, and, so,
$\psi$ can be represented in the form
$$\psi\,(z)=\alpha\,(z)\,\psi_D+\beta\,(z)\psi_B,$$ where the
coefficients $\alpha$ and $\beta$  are entire and $h$-periodic,
$$\alpha\,(z)=\frac1w\,\{\psi\,(z),\psi_B(z)\}, \quad
   \beta\,(z)=\frac1w\,\{\psi_D(z),\psi\,(z)\}.$$
Since all the three solutions are minimal solutions corresponding
to one and the same pair of canonical bases, $\alpha$ and $\beta$
are constant. Let $C\,(z)$ and $D\,(z)$ be the periodic
coefficients from the representations \eqref{1.19a} for the
solution $\psi$, and let $C^{(D)}\,(z)$ and $D^{(D)}\,(z)$ be the
ones for the solution $\psi_D$. As in the proof of Lemma 6.1, we
check that
\begin{equation} \beta\,(z)=
\frac{w_g}w\,(C^{(D)}\,(z)\,D\,(z)-C\,(z)\,D^{(D)}(z)),\label{6.4}
\end{equation} in a vicinity of $-i\infty$.
As $D(z)$ and $D^{(D)}(z)$ tend to zero as $z\to-i\infty$, formula
\eqref{6.4} implies that $\beta\to0$ as $z\to-i\infty$, and, thus,
$\beta\equiv 0$. So, the minimal solution $\psi_D$ is unique up to
a constant factor. In the same way, one proves the same statement
for $\psi_B$.

\qed
\subsection{ Monodromy matrices corresponding to
the minimal entire solutions} We begin this subsection by proving
Theorem 1.6.

\smallpagebreak Let $$w=\{\psi_D,\psi_B\}.$$ By the hypothesis of
the theorem the wronskian $w$ is nonzero. It is given by formula
\eqref{6.3}.

\smallpagebreak The monodromy matrix corresponding to $\psi_D$ and
$\psi_B$ is defined by
$$\left(\psi_D(z+2\pi),\psi_B(z+2\pi)\right)=
\left(\psi_D(z),\psi_B(z)\right)\, {\mathcal M}^T(z).$$ The
coefficients of the monodromy matrix admit the representations:
$${\mathcal M}_{11}(z)=\frac
1w\,\{\psi_D(z+2\pi),\,\psi_B(z)\},\quad {\mathcal
M}_{12}(z)=\frac 1w\,\{\psi_D(z),\,\psi_D(z+2\pi)\},$$
\begin{equation} {}\label{6.5}\end{equation} $${\mathcal M}_{21}(z)=\frac
1w\,\{\psi_B(z+2\pi),\,\psi_B(z)\},\quad {\mathcal
M}_{22}(z)=\frac 1w\,\{\psi_D(z),\,\psi_B(z+2\pi)\}.$$

\smallpagebreak As in the proof of Lemma 6.1, one can easily
calculate all the wronskians in \eqref{6.5} using the canonical
bases $f_{1,2}$ and $g_{1,2}$. Consider, for example, the
coefficient ${\mathcal M}_{11}$. By means of the representation
\eqref{1.19}, we get
\begin{equation} \hskip-5cm \dsize {\mathcal M}_{11}=\frac 1w\,\left(
A^{(D)}(z+2\pi)\,A^{(B)}(z)\,\{f_1(z+2\pi),\,f_1(z)\}+\right.\label{6.6}\end{equation}
$$+A^{(D)}(z+2\pi)\,B^{(B)}(z)\,\{f_1(z+2\pi),\,f_2(z)\}+
B^{(D)}(z+2\pi)\,A^{(B)}(z)\,\{f_2(z+2\pi),\,f_1(z)\}+$$
$$\hskip5cm +\left.
B^{(D)}(z+2\pi)\,B^{(B)}(z)\,\{f_2(z+2\pi),\,f_2(z)\}\right).$$
Now, we recall that $f_{1,2}$ and $g_{1,2}$ are Bloch solutions,
$$f_{1,2}(z+2\pi)=\alpha_{1,2}(z)\, f_{1,2}(z),\quad
g_{1,2}(z+2\pi)=\beta_{1,2}(z)\, g_{1,2}(z),$$ where
$\alpha_{1,2}$ and $\beta_{1,2}$ are $h$-periodic functions. This
allows to continue the calculation began in \eqref{6.6}: $$\dsize
{\mathcal M}_{11}=\frac {w_f}w\,\left(
\alpha_{1}(z)\,A^{(D)}(z+2\pi)\,B^{(B)}(z)-
\alpha_{2}(z)\,B^{(D)}(z+2\pi)\,A^{(B)}(z) \right).$$

\smallpagebreak The last formula allows to get the asymptotics of
${\mathcal M}_{11}$ for $z\to+i\infty$. Using  the representation
\eqref{1.16} and  recalling that $n_+(v)=n$, we get $${\mathcal
M}_{11}=-\alpha_{2}^{0}\,\frac{w_f}w\,A_B\,B_D\,
e^{\dsize-\frac{2\pi i n}h\,z}\,(1+o\,(1)),\quad z\to+i\infty. $$
Here, $A_B$ and $B_D$ are the asymptotic coefficients of $\psi_B$
and $\psi_D$. Now, using the formula \eqref{6.3} for $w$, we get
finally: \begin{equation} {\mathcal M}_{11}=
\alpha_{2}^{0}\,e^{\dsize-\frac{2\pi i n}h\,z}\,(1+o\,(1)),\quad
z\to+i\infty. \label{6.7}\end{equation} Remind that the constant
$\alpha_{2}^{0}$ from \eqref{1.16} is nonzero.

\smallpagebreak Similarly, one can obtain \begin{equation}
{\mathcal M}_{11}=\beta_{1}^0\,e^{\dsize+\frac{2\pi i n}h\,z}\,
(1+o\,(1)),\quad z\to-i\infty,\label{6.8}\end{equation} with the
nonzero constant $\beta_{1}^0$ from \eqref{1.17}. Formulae
\eqref{6.7} -- \eqref{6.8} imply that the entire $h$-periodic
function ${\mathcal M}_{11}$ is a trigonometric polynomial.
Considering it as function of the variable $
z_1=\frac{2\pi}{h}\,z$, we see that
\begin{equation} n_\pm({\mathcal M}_{11})=n.\label{6.9}\end{equation}

\smallpagebreak Formulae \eqref{6.7} -- \eqref{6.8} are, in fact,
the first two of the formulae \eqref{1.24} -- \eqref{1.31} for the
coefficients of the monodromy matrix corresponding to the
solutions $\psi_D$ and $\psi_B$. In the same way, one investigates
all the other coefficients of the monodromy matrix. This leads to
the formulae \eqref{1.24} -- \eqref{1.31}.  For $\mathcal M$
considered as a function of $z_1=\frac{2\pi}h\,z$, they obviously
imply that $$ n_-({\mathcal M}_{12})\le n-1,\quad n_+({\mathcal
M}_{12})\le n,$$ $$ n_-({\mathcal M}_{21})\le n,\quad
n_+({\mathcal M}_{21})\le n-1,$$ $$ n_\pm({\mathcal M}_{22})\le
n-1.$$ These inequalities, and formulae \eqref{6.9} mean that the
monodromy matrix, as a function of the variable $
z_1=\frac{2\pi}{h}\,z$, belongs to $\Omega\,(n)$. This completes
the proof of the theorem.

\qed

\medpagebreak We finish this section by formulating one more
theorem generalizing Theorem 1.6. As we have seen, any two of the
minimal solutions form a basis in the space of entire solutions of
\eqref{1.1}. We call these bases {\it natural}. One has

\medpagebreak {\bf Theorem  6.3.} \ {\it Let, the matrix $M$
satisfy condition \eqref{1.20}. For any natural basis, the
coefficients of the corresponding  monodromy matrix ${\mathcal M}$
are trigonometric polynomials of the variable $$
z_1=\frac{2\pi}{h}\,z,$$ and for any of these polynomials the
numbers $n_\pm$ satisfy the inequality $$-n\le n_\pm\le n.$$}

\medpagebreak
The proof of this theorem is similar to the proof of Theorem 1.6.
It can be said that
Theorem 6.3 describes the characteristic property of the minimal
entire solutions.
\subsection{ Canonical factorizations of the monodromy
matrices} For $M\in \Omega\,(n)$, \ $n>1$, one can reconstruct in
terms of the asymptotic coefficients of the minimal solutions only
a part of the constant coefficients of the monodromy matrix. For
example, formulae \eqref{1.24} -- \eqref{1.25} allow to recover
only the coefficients $\left({\mathcal M}_{11}\right)_{\pm n}$ of
the trigonometric polynomial ${\mathcal M}_{11}$, $$\hskip -2cm
{\mathcal M}_{11}(z)=\left({\mathcal
M}_{11}\right)_{-n}\,e^{\dsize -2\pi inz/h}+ \left({\mathcal
M}_{11}\right)_{-n+1}\,e^{\dsize -2\pi i(n-1)z/h}+ $$ $$\hskip 8cm
+\,\,\dots\,\,+\left({\mathcal M}_{11}\right)_{n}\,e^{\dsize 2\pi
inz/h}.$$ Trying to get an efficient description of the other
coefficients, we come to the following construction.

\medpagebreak
Let $f_{1,2}$ and $g_{1,2}$ be two consistent canonical bases with
the parameters $\phi_\pm$.
Consider canonical bases $f_{1,2}^{(j)}$ and $g_{1,2}^{(j)}$ with the
parameters equal to $\phi_\pm+2\pi j$, \ $j=0,\,1,\,2,\,\dots n$.
Note that all the pairs $f_{1,2}^{(j)}$ and $g_{1,2}^{(j)}$ are consistent.
By definition,
$$f_{1,2}^{(0)}=f_{1,2},\quad g_{1,2}^{(0)}=g_{1,2},$$
In view of the analysis of section 2.5.2, we can assume that
$$f_{1,2}^{(n)}(z)=s_+f_{1,2}(z+2\pi),\quad
g_{1,2}^{(n)}(z)=s_-g_{1,2}(z+2\pi),$$
where $s_\pm$ are two numbers, each of them can be equal either to $+1$ or
to $-1$. For the sake of definitiness, we consider the case where $s_+=s_-=1$.

\smallpagebreak
Denote by $\psi_D^{(j)}$ and $\psi_B^{(j)}$ the minimal solutions corresponding
to the canonical bases $f_{1,2}^{(j)}$ and $g_{1,2}^{(j)}$.  Let
$$\psi_D=\psi_D^{(0)},\quad \psi_B=\psi_B^{(0)}.$$
In view of Lemma 6.2, we can assume that
$$\psi_D^{(n)}(z)=\psi_D(z+2\pi),\quad
\psi_B^{(n)}(z)=\psi_B(z+2\pi).$$

\smallpagebreak
Let, for any $j=0,\,1,\,2,\,\dots\,n-1$, \ the solutions
$\psi_D^{(j)}$ and $\psi_B^{(j)}$ be linearly independent over the ring of
$h$-periodic functions.

\smallpagebreak Denote by  $\Psi^{(j)}$ the $2\times2$-matrix
composed of the vectors $\psi_D^{(j)}$ and $\psi_B^{(j)}$. Any of
the pairs $\psi_D^{(j)}$ and $\psi_B^{(j)}$ is a basis in the
space of entire solutions of \eqref{1.1}. Therefore, one can
define the transition matrices $S_j(z)$ relating $\Psi^{(j)}$ to
$\Psi^{(j-1)}$, $$\Psi^{(j)}(z)=\Psi^{(j-1)}(z)\,S_j^t(z),$$ where
"${}^t\,\,$" denotes the transposition.

\smallpagebreak The monodromy matrix corresponding to the basis
$\psi_D$, \ $\psi_B$ is represented in the form $${\mathcal
M}(z)=S_{n-1}(z)\,S_{n-2}(z)\,\dots\,S_{1}(z).$$ We call this
factorization of the monodromy matrix {\it canonical}. The main
feature of this factorization is related to the standard
(canonical) form of the transition matrices $S_j$. One has

\medpagebreak
{\bf Theorem 6.4.} \ {\it
Any of the transitions matrices is an $h$-periodic trigonometric
polynomial of the form
$$\left(\begin{array}{cc} a_1 e^{2\pi iz/h}+a_0 +a_{-1} e^{-2\pi iz/h}&
b_0+b_{-1} e^{-2\pi iz/h}\\{}\\
c_0+c_{-1} e^{2\pi iz/h} & d_0\end{array}\right),$$
where $a_{\pm1,0}$, $b_{-1,0}$, $c_{1,0}$ and $d_0$ are
constant coefficients (depending on $j$), and $a_\pm\ne 0$.
The $\det S_j$ are nonzero and independent of $z$.}

\smallpagebreak The theorem shows that any of the transition
matrices coincides up to a constant factor with a matrix from
$\Omega(1)$. The proof of this theorem is similar to one of
Theorem 1.6. The constant coefficients of $S_j$ can be expressed
in terms of the asymptotic coefficients of the minimal solutions
$\psi_D^{(j)}$, \ $\psi_B^{(j)}$, and  $\psi_D^{(j-1)}$, \
$\psi_B^{(j-1)}$ by  formulae similar to \eqref{1.24} --
\eqref{1.31}.
\section{ One dimensional difference Schr\"odinger equations}
 Consider  the difference Schr\"odinger equation
\begin{equation} \frac{f\,(z+h)+f\,(z-h)}2+v\,(z)\,f\,(z)=E\,f\,(z),\label{7.1}
\end{equation} where $v$ is a given trigonometric polynomial, and $E$ is a
spectral parameter. It is equivalent to equation \eqref{1.1} with
the matrix \begin{equation} M=\left(\begin{array}{cc}
                           2E-2v\,(z) & -1\\
                              1       & 0 \end{array}\right).\label{7.2}\end{equation}
Assume that $n_+(v)=n_-(v)=n\in \N$. Then, this matrix belongs to
the set $\Omega_{nn}$. Thus, one can apply  Theorem 1.6 to
investigate the spectrum of \eqref{7.1} by means of the
monodromization procedure.

\smallpagebreak In this section, we concentrate on the Harper
equation \begin{equation} \frac{f\,(x+h)+f\,(x-h)}2+\lambda\cos
z\,f\,(z)=E\,f\,(z),\label{7.3}\end{equation} in which, $\lambda$
is a fixed positive parameter. Let us study for this equation the
monodromy matrix described in Theorem 1.6 in more detail.
\subsection{ Monodromy matrices for Harper equation} \ {\bf 1.} \ First,
discuss the choice of the bases
$f_{1,2}$ and $g_{1,2}$. The formulae \eqref{1.11} and
\eqref{1.14} allow to choose the corresponding parameters
$\phi_\pm$ so that
\begin{equation} \phi_-=-i\xi-\pi,\quad \phi_+=i\xi-\pi,\quad \xi=\ln\lambda \in
\R.\label{7.4}\end{equation} Since $\xi\in\R$, these canonical
bases are consistent, and there exist all the four minimal entire
solutions.

\smallpagebreak
{\bf 2.} \ Consider the minimal solutions $\psi_D$ and $\psi_B$ corresponding
to these canonical bases. Assume that these solutions
are linearly independent.

\smallpagebreak
The solutions $\psi_D$ and $\psi_B$
are defined up to independent of $z$ factors.
Describe a convenient choice of these factors. Since the wronskian
$\{\psi_D,\,\psi_B\}\ne 0$, the asymptotic coefficient $C_D$ is nonzero,
see Proposition 1.3.
Normalize the solution $\psi_D$ by dividing it by its asymptotic coefficient
$C_D$. We use for  the new minimal solution and its asymptotic
coefficients
the old notations. In particular, now, $C_D\equiv 1$.

\smallpagebreak {\bf 3.} \ To fix uniquely the second minimal
solution, we note that the matrix \eqref{7.2} with $v=\cos z$
satisfies the relation \begin{equation}
M\,(2\pi-z)=\sigma\,M^{-1}(z)\sigma, \quad
\sigma=\left(\begin{array}{cc} 0 & 1\\ 1  & 0
\end{array}\right).\label{7.5}\end{equation} This implies that the
vector
\begin{equation} \psi\,(z)=\sigma\,\psi_D(2\pi-z+h)\label{7.6}\end{equation}
satisfies \eqref{1.1} together with $\psi_D$. Show that, up to a
constant factor, $\psi$ coincides with $\psi_B$. Begin with

\medpagebreak {\bf Lemma 7.1} \ {\it For $z$ being in a vicinity
of $-i\infty$, $$\sigma\,f_1(2\pi-z+h)= -g_1(z)\,c_1(z),\quad
\sigma\,f_2(2\pi-z+h)= -g_2(z)\,c_2(z),$$ where $c_1$ and $c_2$
are  analytic $h$-periodic functions tending to  $1$ as $z\to\pm
i\infty$.}

\demo Let us prove the first of the above relations. Note that,
for the matrix \eqref{7.2} corresponding to Harper equation, the
general asymptotic formulae for $f_{1}$  and $g_1$ take the form
\begin{equation}
f_1(z)= e^{\dsize +\frac i{2h}\,(z-\pi+i\xi)^2+\frac{iz}2}\,
\left(\begin{array}{c} 1+o\,(1)\\ -\lambda\, e^{-iz}\,(1+o\,(1))
\end{array}\right),\label{7.7}\end{equation}

$$\im z\to+\infty, $$
and
\begin{equation}
g_1(z)= e^{\dsize +\frac i{2h}\,(z-\pi-i\xi)^2-\frac{iz}2}\,
\left(\begin{array}{c} 1+o\,(1)\\ -\frac1{\lambda}\, e^{-iz+ih}\,(1+o\,(1))
\end{array}\right),\label{7.8}
\end{equation}

$$\quad \im z \to-\infty.$$ Substituting \eqref{7.7} in the
expression $-\sigma\,f_1(2\pi-z+h)$, we check that it does have
the representation \eqref{7.8}. Moreover, one can easily see that
if $f_1$ is a Bloch solution, then $\sigma\,f_1(2\pi-z+h)$ is also
a Bloch solution of \eqref{1.1}. Now, Lemma 2.6 implies the
desired result. The second relation can be derived similarly.

\qed

\smallpagebreak
This lemma immediately leads to

\medpagebreak
{\bf Lemma 7.2.} \ {\it The solution
$\psi$ coincide with $\psi_B$ up to a constant factor.}

\medpagebreak \demo The representations \eqref{1.19} for $\psi_D$
imply that, in a vicinity $\C_-$ of $-i\infty$,
$$\psi\,(z)=A^{(D)}(2\pi-z+h)\,\sigma\,f_1(2\pi-z+h)+
B^{(D)}(2\pi-z+h)\,\sigma\,f_2(2\pi-z+h).$$ Therefore, in view if
Lemma 7.1,
\begin{equation} \psi\,(z)=C\,(z)\,g_1(z)+D\,(z)\,g_2(z),\label{7.9}\end{equation}
where $$C\,(z)=-c_1(z)\,A^{(D)}(2\pi-z+h),\quad
D\,(z)=-c_2(z)\,B^{(D)}(2\pi-z+h).$$ Similarly, one shows that in
a vicinity $\C_+$ of $+i\infty$,
$$\psi\,(z)=A\,(z)\,f_1(z)+B\,(z)\,f_2(z),$$ where
\begin{equation} A\,(z)=-c_1^{-1}(2\pi-z+h)\,C^{(D)}(2\pi-z+h),\quad
B\,(z)=-c_2^{-1}(2\pi-z+h)\,D^{(D)}(2\pi-z+h).\label{7.10}\end{equation}
These representations show that $\psi$ is a minimal solution for
which $B\,(z)\to0$ as $z\to+i\infty$. Therefore, by Corollary 1.5,
it coincides with $\psi_B$ up to a constant factor.

\qed

\medpagebreak {\bf 4.} \ In the sequel, we shall normalize the
solution $\psi_B$ so that
\begin{equation} \psi_B(z)=\sigma\,\psi_D(2\pi-z+h).\label{7.11}\end{equation}
Note that, in this case,  \eqref{7.9} and \eqref{7.10} imply the
relations
\begin{equation} A_B=-C_D=-1,\quad B_B=-D_D\,e^{-4\pi^2i/h},\quad C_B=-A_D,\quad
D_B=-B_D.\label{7.12}\end{equation}

\smallpagebreak
{\bf 5.} \ Discuss now the monodromy matrix corresponding
the chosen $\psi_D$ and $\psi_B$.

\medpagebreak {\bf Theorem 7.3.} {\it The constructed minimal
entire solutions $\psi_D$ and $\psi_B$ and their asymptotic
coefficients are meromorphic in $E$,
$$\{\psi_D,\,\psi_B\}\not\equiv 0,$$ and the corresponding
monodromy matrix has the form \begin{equation} {\mathcal
M}=\left(\begin{array}{cc} a -2\lambda_1\cos(2\pi z/h) &
s+t\,e^{\dsize-2\pi iz/h}\\{}\\ -s-t\,e^{\dsize 2\pi iz/h}       &
\frac1{\lambda_1}\,st
\end{array}\right), \label{7.13}\end{equation} where
$$a=\lambda_1\,\frac{1-s^2-t^2}{st},\quad\quad
\lambda_1=\lambda^{2\pi/h},$$ and
$$s=-\lambda_1\frac{D_D}{B_D},\quad\quad t= -\lambda_1\,A_D.$$ }

\demo First, prove that $\psi_D$ and $\psi_B$ and their asymptotic
coefficients are meromorphic in $E$. Consider the solution
$\psi_D$ before its normalization by the condition $C_D=1$. It can
be constructed by means of the integral equation \eqref{4.4}. For
the matrix \eqref{7.2} with $v\,(z)=\cos z$, this equation takes
the form $$\psi\,(z)=m\,(z)+\, E\,\tilde K \psi\,(z) $$ where $K$
is the integral operator with the kernel $$\kappa\,(z,\,\zeta)=
\frac 1{ih}\,\theta\,(z,\,\zeta)\, \frac{
        [\,m(z)\,\tilde m\,(\zeta)-
           m\,(\zeta)\,\tilde m\,(z)\,]}
        {\{m,\,\tilde m\}}.$$
This equation describes the first component of the vector $\psi_D$.
The integral operator is  compact in the suitable function space,
and the solution  $\psi_D$ together with $\psi$ appears to be meromorphic
in $E$.

\smallpagebreak
Furthemore,
the solutions $f_{1,2}$ and $g_{1,2}$, being constructed as in Section 2,
are entire in $E$.

\smallpagebreak
Finally, the asymptotic coefficients of $\psi$ can be expressed in terms
of the
wronskians of the canonical basis solutions $f_{1,2}$ and $g_{1,2}$, and of
$\psi$ itself,
for example, $C$ is the zeroth Fourier coefficient of the function
$\frac{\{\psi^{(1)},\,g_2\}}{\{g_1,\,g_2\}}$.
Thus, the asymptotic coefficients are meromorphic in $E$.

\smallpagebreak Note that we have already calculated these
asymptotic coefficients for $E=0$ in Section 3: formulae
\eqref{3.21} -- \eqref{3.24}  imply that these meromorphic
functions are not identically zero.

\smallpagebreak All this shows that the normalized minimal
solution $\psi_D$, the minimal solution $\psi_B$ related to it by
\eqref{7.11}, and the asymptotic coefficients of these solutions
are meromorphic in $E$. Moreover, since the asymptotic
coefficients of $\psi_D$ are nontrivial meromorphic functions, the
wronskian $\{\psi_D,\,\psi_B\}$ is not identically zero.

\medpagebreak Now obtain the representation \eqref{7.13} for the
monodromy matrix corresponding to the chosen minimal solutions
$\psi_D$ and $\psi_B$. First, consider the coefficient ${\mathcal
M}_{11}$. By Theorem 1.6, it is a trigonometric polynomial of the
form $$a_{-1}\,e^{\dsize -2\pi i z/h}+ a_0+a_{1}\,e^{\dsize 2\pi i
z/h}.$$ Formulae \eqref{1.24} -- \eqref{1.25} imply that
$$a_{-1}=\alpha_2^0,\quad a_1=\beta_1^0,$$ where $\alpha_{2}^0$
and $\beta_{1}^0$ are the first Fourier coefficients  of the Bloch
multipliers $\alpha_{2}$ and $\beta_{1}$ of the Bloch solutions
$f_{2}$ and $g_{1}$, $$f_{2}(z+2\pi)=\alpha_{2}(z)\,
f_{2}(z),\quad g_{1}(z+2\pi)=\beta_{1}(z)\, g_{1}(z).$$ The
solution $g_1$ admits the representation \eqref{7.8} which implies
that $\beta_{1}^0=-\lambda^{\dsize 2\pi/h}=-\lambda_1$. Similarly,
$\alpha_{2}^0=-\lambda_1$, and, thus, $${\mathcal
M}_{11}(z)=a_0-2\lambda_1\cos\,(2\pi z/h).$$

\smallpagebreak Formulae \eqref{1.26} -- \eqref{1.31} and
relations \eqref{7.12} imply the desired representations for
${\mathcal M}_{12}$, \ ${\mathcal M}_{21}$, and ${\mathcal
M}_{22}$. Formula for the coefficient $a_0$ follows from these
representations and the equality $\det {\mathcal M}\equiv 1$. \qed
\subsection{ Family of matrices generated by Harper equation}
Fix a positive $\lambda$. Define  a family $\HI\,(\lambda)\subset
\Omega\,(1)$ of  matrix functions. This family consists of a two
dimensional manifold $\HI^0$ and four one dimensional linear
manifolds $h_0^\pm$ and $h_1^\pm$. The $\HI^0$ is the set of the
matrices ${\mathcal M}(.,\,\lambda,\,w)$ defined by
\begin{equation} {\mathcal M}(z,\,\lambda,\,w)=\left(\begin{array}{cc} a
-2\lambda\,\cos z & s+t\,e^{\dsize-iz}\\{}\\ -s-t\,e^{iz}       &
\frac1{\lambda}\,st \end{array}\right), \label{7.14}\end{equation}
where $$a=\lambda\,\frac{1-s^2-t^2}{st},$$

$$w=(s,t)\in \C^2,\quad s,t\ne 0.$$ The $h_0^\pm$ is the set of
the matrices $$M_0^\pm(z,\,\lambda,\,a)=\left(\begin{array}{cc} a
-2\lambda\,\cos z & \pm 1 \\{}\\ \mp 1      & 0
\end{array}\right),\quad a\in \C, $$ and $h_1^\pm$ is the set of
the matrices $$M_1^\pm(z,\,\lambda,\,a)=\left(\begin{array}{cc} a
-2\lambda\,\cos z & \pm e^{-iz} \\{}\\ \mp e^{iz}      & 0
\end{array}\right),\quad a\in \C. $$

\smallpagebreak Apply Theorem 1.6 to equation \eqref{1.1} with a
matrix $M\in \HI$. Choose the bases $f_{1,2}$ and $g_{1,2}$ so
that $$\phi_-=-i\xi-\pi-\frac h2\,n_-(b),\quad
\phi_+=i\xi-\pi-\frac h2\,n_+(b),\quad \xi=\ln\lambda \in \R.$$
These canonical bases are consistent, and there exist all the four
minimal entire solutions. Assume that the asymptotic coefficient
$C_D$ is nonzero.

\smallpagebreak We normalize the  solution $\psi_D$ by the
condition $C_D=1$ and define $\psi_B$ by formula \eqref{7.11}.
Since the matrix $M$ satisfies the relation \eqref{7.5}, $\psi_B$
is really the desired minimal solution.

\smallpagebreak
Assume that $\psi_D$ and $\psi_B$ are linearly independent over the
ring of $h$-periodic functions. Instead of Theorem 7.3, one can prove

\medpagebreak {\bf Theorem 7.4.} {\it As a function of the
variable $2\pi z/h$, the  monodromy matrix ${\mathcal M}$
corresponding to the constructed minimal entire solutions belongs
to $\HI\,(\lambda_1)$ with $$\lambda_1=\lambda^{\dsize 2\pi/h}.$$
In particular, if ${\mathcal M}\in\HI^0$, then it has the form
$${\mathcal M}(2\pi z/h,\,\lambda_1,\,w_1),\quad w_1=(s_1, t_1),$$
$$ s_1=-\left(-i\frac{\sqrt{\lambda}}s\,e^{ih/8}\right)\,
\lambda_1\frac{D_D}{B_D},\quad\quad
t_1=-\left(-i\frac{\sqrt{\lambda}}t\,e^{ih/8}\right)\,
\lambda_1\,\frac{A_D}{C_D}.$$ }
\bibliographystyle{plain}

\end{document}